\documentclass[iop]{emurateapj}

\usepackage{fancybox}
\usepackage{ulem}
\usepackage{ascmac}
\usepackage{bm}

\newcommand{\range}{\mbox{\scriptsize --}}
\newcommand{\p}{Pop III}
\newcommand{\npopp}{n_{\rm pop3}}
\newcommand{\npop}{$\npopp$}
\newcommand{\nsurv}{$N_{\rm surv}$}
\newcommand{\Msun}{M_\odot}
\def\gsim{\mathrel{\rlap{\lower 4pt \hbox{\hskip 1pt $\sim$}}\raise 1pt \hbox {$>$}}}
\def\lsim{\mathrel{\rlap{\lower 4pt \hbox{\hskip 1pt $\sim$}}\raise 1pt \hbox {$<$}}}

\newcommand{\hMpc}{$ \, h^{-1} \rm Mpc$}
\newcommand{\hpc}{$ \, h^{-1} \rm pc$}
\newcommand{\hMsun}{$\, h^{-1} M_{\odot}$}

\shorttitle{Where are the Low-Mass Population III Stars ?}
\shortauthors{Ishiyama et al.}

\begin{document}

\title{Where are the Low-Mass Population III Stars ?}
\author{Tomoaki Ishiyama$^1$}
\author{Kae Sudo$^2$}
\author{Shingo Yokoi$^2$}
\author{Kenji Hasegawa$^3$}
\author{Nozomu Tominaga$^{2,4}$}
\author{Hajime Susa$^{2}$}
\affil{Institute of Management and Information Technologies, Chiba University, 1-33, Yayoi-cho, Inage-ku, Chiba, 263-8522, Japan\altaffilmark{1}}
\affil{Department of Physics, Konan University, Okamoto, Kobe, Japan\altaffilmark{2}}
\affil{Graduate School of Science, Nagoya University, Furo-cho, Chikusa-ku, Nagoya, Aichi 464-8602, Japan\altaffilmark{3}}
\affil{Kavli Institute for the Physics and Mathematics of the Universe (WPI),\\
  The University of Tokyo, 5-1-5 Kashiwanoha, Kashiwa, Chiba 277-8583, Japan\altaffilmark{4}}
\email{ishiyama@chiba-u.jp}
\begin{abstract}
We study the number and the distribution of low mass \p\ stars in the
Milky Way.  In our numerical model, hierarchical formation of dark
matter minihalos and Milky Way sized halos are followed by a high
resolution cosmological simulation.
We model the
\p\ formation in H$_2$ cooling minihalos without metal under UV
radiation of the Lyman-Werner bands.  Assuming a Kroupa IMF from 0.15
to 1.0 $\Msun$ for low mass \p\ stars, as a working hypothesis, we try
to constrain the theoretical models in reverse by current and future
observations.  
We find that 
the survivors tend to concentrate on the center of halo and subhalos.  
We also evaluate the observability of \p\ survivors in the Milky Way and
dwarf galaxies, and constraints on the number of \p\ survivors per
minihalo.  The higher latitude fields require lower sample sizes
because of the high number density of stars in the galactic disk, the
required sample sizes are comparable in the high and middle latitude
fields by photometrically selecting low metallicity stars with
optimized narrow band filters, and the required number of dwarf
galaxies to find one \p\ survivor is less than ten at $<100$ kpc
for the tip of red giant stars.  Provided that available observations
have not detected any survivors, the formation models of low mass 
\p\ stars with more than ten stars per minihalo are already excluded.
Furthermore, we discuss the way to constrain the IMF
of \p\ star at a high mass range of $\ga 10\Msun$.
\end{abstract}
\keywords{
methods: numerical
---early Universe
---first stars
---stars: low-mass
---Galaxy: structure
---dark matter
}


\section{Introduction}
First stars are born in the minihalos of $\sim10^5 - 10^6M_\odot$
approximately one hundred million years after the big bang 
\citep{Haiman1996,Tegmark1997,Nishi1999,Fuller2000,Abel2002,Bromm2002,Yoshida2003}.
In such environments, the H$_2$ molecule is the only coolant of the gas
different from the nearby interstellar medium (ISM) which contains
metals and dust grains.
The mass of the first stars are expected to be more massive than
their present-day counterparts because of the inefficient cooling via H$_2$.
In fact, detailed studies, including cosmological radiation hydrodynamical
simulations have shown that 
the first stars are very massive as a first approximation
\citep{Omukai1998,Omukai2001,Omukai2003,Bromm2004,Yoshida2006,
Yoshida2008,Hosokawa2011,Hirano2014,Hirano2015}.

However, the inefficiency of the cooling is also conducive to the
formation of heavy circumstellar disks which are gravitationally
unstable. The unstable disks fragment into small pieces, which could
end up as low mass stars
\citep{Clark2008,Smith2011,Clark2011b,Clark2011,Greif2011,Greif2012,Machida2013,Susa2013,Susa2014}.
If those low mass stars are less massive than 0.8$M_\odot$, their life
times are longer than the age of the universe, and thus they could survive
to be found in the present-day universe.  However, the fates of the
fragments are still uncertain theoretically. On the one hand, they
could fall onto the central protostar and merge with it because of the
efficient transportation of angular momentum in the disk
\citep[e.g.,][]{Vorobyov2015, Hosokawa2015, Sakurai2015}, but at the
same time they could be ejected from the central dense region to
highly eccentric orbits via many body gravitational interactions 
\citep[e.g.,][]{Susa2014}.  In
the former case, they cannot be the low mass stars, but in the latter
case, the fragments could be long-lived low mass stars because of the
poor mass accretion rate in such orbits.

Hence, the fraction of the fragments that survives as low mass stars and thus
the average number of such low mass stars per minihalo is highly
controversial, and it remains an open question.

The most straightforward way to approach this issue is to
search the metal free stars in the Milky Way. In fact, there is a
long history of the hunting for those stars
\cite[][and the references are therein]{Beers2005}. 
Recent surveys search the metal poor stars among
$10^5-10^6$ stars
\citep[e.g.][]{Keller2007, Yanny2009, Li2015}, and 
many iron poor stars are found to be carbon enhanced
metal poor stars (CEMPs). Some stars such as SDSS J102915+172927
\citep{Caffau2011} are totally deficient in metals including carbon.
However, no metal free star has been discovered so far.

The current state of observations of first star hunting implies that
the surviving first stars are pretty rare in the Milky Way even if they exist.
However, this fact only reveals that they are rare compared to other
normal stars formed in the later different environments. We need to
predict theoretically how many first stars should be found in the Milky Way
assuming a certain model of star formation, or initial mass function (IMF) of
those stars, in order to constrain the theory by the observations.
At the same time, it is helpful to suggest the area in the
sky where we should survey to find low mass first stars.

Several theoretical studies have been conducted to investigate this issue.
\citet{Diemand2005b} discussed whether stars should have a centrally condensed
distribution in our Galactic halo than the dark matter mass distribution 
by analyzing the density peaks in their $N$-body simulations.
On the other hand, \citet{Scannapieco2006} also tried to address the issue
by focusing on the position of the metal free stars born in rather massive
halos with virial temperatures of $T_{\rm vir}\ga 10^{4}$K. They
found that these stars should be distributed more smoothly in our Galactic
halo, probably because these massive halos collapse at later epochs than
the minihalos of $\sim 10^6M_\odot$. \citet{Tumlinson2010} investigated
the spatial distribution of metal poor stars in the Milky Way, utilizing 
massive cosmological $N$-body simulation, and they found that the metal poor
stars would be better found in the direction of the Galactic bulge.
However, the mass resolutions of the $N$-body simulations used in these works
were not enough to capture the formation of the minihalos of $\sim 10^6M_\odot$.

Recently, \cite{Hartwig2015} discussed this issue by semi-analytic methods,
and they give the expected number of first stars in our Galactic halo.
They report that unbiased survey of $4\times 10^6$ halo stars could
impose a rather stringent 
constraint on the low mass end of the IMF. However, the spatial
distribution of such stars cannot be derived because of the nature of
semi-analytic methods.

In this paper, we perform a huge cosmological $N$-body simulation that
resolve the minihalos and contains a few Milky Way sized halos in the
simulation box. Assuming the first star formation model in minihalos, 
we can trace the stars until present-day universe and
can give predictions for the number, location, and the apparent
magnitudes of such stars.
Compared with the past/ongoing/planning surveys of metal poor stars, we try to
constrain the theoretical IMF of the first stars at the low mass end.

In \S 2, we describe the numerical model of the cosmological
simulation. The number and the distribution of surviving
first stars in the Milky Way are shown in \S 3. In \S 4, we show the number of the
observable first stars and the probability distribution of finding those
stars on the celestial sphere, assuming a galactic model. 
Then we derive the current constraint on the \p\  star formation model. 
\S 5 is devoted to discussion and \S 6 to summary.


\section{Model Description}

Hierarchical formation of dark matter minihalos and Milky Way sized
halos are followed by a high resolution cosmological simulation.  The
formation of \p\ stars is simply modeled on merger trees of dark
matter halos.

\subsection{Cosmological Simulation}
The cosmological simulation consists of $2048^3$ dark matter particles
in a comoving box of 8\hMpc.
The mass resolution is $5.13 \times 10^3$\hMsun\
and the gravitational softening length is 120\hpc, 
which enable us to handle minihalos with sufficient resolution. 
We generated the initial condition by a publicly available code,
2LPTic\footnote{http://cosmo.nyu.edu/roman/2LPT/}, using second-order
Lagrangian perturbation theory \citep[e.g.,][]{Crocce2006}.  We used the
online version\footnote
{http://lambda.gsfc.nasa.gov/toolbox/tb\_camb\_form.cfm} of CAMB
\citep{Lewis2000} to calculate the transfer function.  The cosmological
parameters adopted are consistent with an observation of the cosmic
microwave background obtained by the {\t Planck} satellite
\citep{Planck2014}, namely, $\Omega_0=0.31$, $\Omega_b=0.048$,
$\lambda_0=0.69$, $h=0.68$, $n_s=0.96$, and $\sigma_8=0.83$.
The initial and final redshifts are 127 and 0.

For the time integration, we used a massively parallel TreePM code, 
GreeM \citep{Ishiyama2009b, Ishiyama2012} with the Phantom-GRAPE 
software accelerator 
\footnote{http://code.google.com/p/phantom-grape/}\citep{Nitadori2006, 
Tanikawa2012, Tanikawa2013}, on Aterui supercomputer at Center for 
Computational Astrophysics, CfCA, of National Astronomical Observatory 
of Japan. 
The snapshots were stored at the redshifts so that the logarithmic interval 
$\Delta \log(1+z)$ is 0.01. 
We carefully checked that how this 
interval affects the quality of merger trees. The value adopted here
yields well converged merger rates. 

To identify halos, we used the Friends-of-Friends (FoF) algorithm 
\citep{Davis1985} with a linking parameter of $b=0.2$. The smallest halo 
consists of 32 particles, which set the minimum FoF halo mass to be $1.6 
\times 10^5$\hMsun. We extracted merger trees by the algorithm described 
in \citet{Ishiyama2015}.

\subsection{Model for the Formation of \p\ Stars}

In our \p\ star formation model, \p\ stars are assumed to form in
minihalos where virial temperature $T_{\rm vir}$ exceeds a threshold
$T^{\rm crit}_{\rm vir}$.
In such halos, H$_2$ molecules form efficiently and the gas collapses to form stars via 
H$_2$ line cooling. We also put an upper bound of 
$T_{\rm vir}$ so that we exclude the atomic cooling halos. 
We set the upper bound to be 2000K in this work.
We calculated $T_{\rm vir}$ of the minihalos using a function proposed by \citet{Kitayama2001}, 
\begin{eqnarray}
T_{\rm vir} = 
9.09 \times 10^3
\left( \frac{\mu}{0.59} \right)
\left[ \frac{M(z_{\rm c})}{10^9 \, h^{-1} M_{\odot}} \right]^{2/3}
\times \nonumber \\ 
\left[ \frac{\Delta_{\rm c} (z_{\rm c})}{18 \pi^2} \right]^{1/3}
(1+z_{\rm c}) \, \rm K,
\end{eqnarray}
where $\mu$ is the mean molecular weight in units of the proton mass, 
$M(z_{\rm c})$ is the minihalo mass at the collapse redshift $z_{\rm c}$, 
and $\Delta_{\rm c} (z_{\rm c})$ is the mean overdensity of 
collapsed halos. 
We set $\mu=1.22$, and used the mean overdensity according to  the spherical collapse model 
\citep{Bryan1998}.
Other values are directly derived from the simulation.

Once \p\ stars form,
UV radiation in the Lyman-Werner (LW) bands from the stars 
causes photodissociations of H$_2$ molecules
and suppresses \p\ star formation in relatively lower mass halos.  We
used the criterion of the virial temperature under the LW background
proposed by \citet{Machacek2001},
\begin{eqnarray}
\left(\frac{T^{\rm crit}_{\rm vir}}{\rm 1000 K} \right) =
0.36 
\left[F_{\rm LW} \,
(\Omega_b h^2)^{-1} \,
\left( \frac{1+z}{20} \right)^{3/2}
\right]^{0.22},
\end{eqnarray} 
,where $F_{\rm LW}$ is the LW flux 
in a unit of $\rm 10^{-21} \, ergs^{-1} \, cm^{-2} \, Hz^{-1}$.    
We assume spatially uniform and time dependent LW flux, \
\begin{eqnarray}
F_{\rm LW} = 4\pi J_{\rm LW} = 1.26 \times 10^{1.8( -1 - \tanh(0.1(z-40)))}, 
\end{eqnarray}
which is a fitting function derived to reproduce cosmological reionization
simulations \citep{Ahn2012}.

By tracking the merger tree in descending redshift, we listed halos
with $T^{\rm crit}_{\rm vir} < T_{\rm vir} < \rm 2000K $ as minihalos
where \p\ stars are assumed to be born with the formation redshift
$z_{\rm c}$.  If any progenitors in the main branch of a halo are
already listed, the halo is excluded from the list. The main branch of
a halo is extracted by connecting the most massive progenitor of the
most massive progenitor.  This procedure ensures that gas in 
\p\ star forming halos is metal free. 
We terminated the list up at $z=10$, when the reionization
is assumed to start,
since the photoheating accompanied by the reionization shut down 
the star formation in such minihalos. 
The upper limit of the virial temperature
criterion is insensitive to the total number of minihalos. When we
change the upper limit to $\rm 8000K$ from $\rm 2000K$, the number of
minihalos increases in only $\sim 1\%$.

\subsection{\p\ Survivors in the Present Universe}\label{sec:surv}

The IMF of \p\ stars is still unknown, although there are some theoretical
implications \citep[e.g.,][]{Greif2011, Hirano2014, Susa2014,
  Hirano2015}.  As discussed in \S 1, the low mass end is particularly
uncertain even in theoretical models. Hence, we simply assume that the
number \npop\ of low-mass \p\ stars with a main-sequence mass of
$0.15\mbox{\scriptsize --}1.0 \Msun$ formed in a minihalo.  Then we
try to constrain the theoretical models in reverse.

We selected randomly \npop\ dark matter particles from
each minihalo as tracers of the low mass \p\ survivors.
The spatial positions of the tracers at $z=0$ are
assumed to be those of \p\ survivors. In this study,
we use $\npopp=1$ and 10. We adopt a Kroupa IMF \citep{Kroupa2001}
for the low-mass \p\ stars and randomly set the mass of
each \p\ star. Their magnitudes of various bands at $z=0$
are calculated from their masses and ages using an isochrone
model with $Z=0$ for stars with $M>0.7M_\odot$ and $Z=0.0001$
for stars with $M<0.7M_\odot$ 
\citep{Girardi2000,Marigo2001}.\footnote{\url{http://pleiadi.pd.astro.it/}}
Since the lifetime of stars with masses larger than $\sim0.8M_\odot$ is
shorter than the cosmic time, these stars cannot survive and the
number of \p\ survivors stars per minihalo is $\sim0.9$ \npop.

\section{\p\ stars in our Galactic Halo}

\begin{figure}
\centering 
\includegraphics[width=9cm]{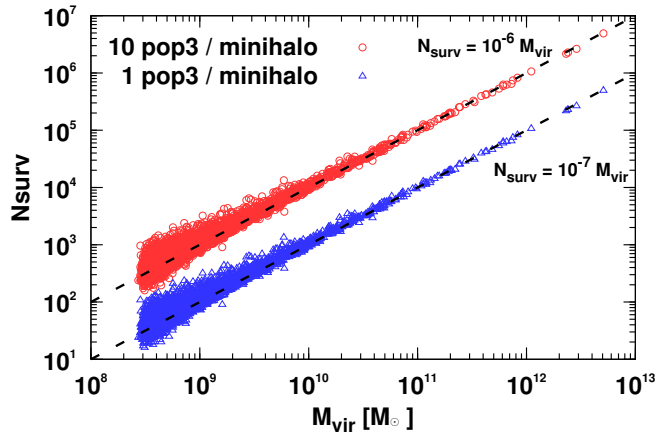} 
\caption{ 
Number of \p\ survivors \nsurv\ at $z=0$ as a function of the halo mass. 
Blue triangles and red circles show the results of 
\npop$=$1 and 10 models.
The thick dashed lines show the best-fitting functions 
(Equation (\ref{eq:nsurv}) in the text).
}
\label{fig:nsurv}
\end{figure}

Figure \ref{fig:nsurv} shows the number of \p\ survivors, 
\nsurv\ in each halo at $z=0$ as a function of the halo mass. 
The number of survivors is proportional to the halo mass 
and the number of \p\ stars per minihalo, \npop. 
The best fitting function is given by 
\begin{eqnarray}
N_{\rm surv} = 10^{-7} \npopp \frac{M_{\rm vir}}{M_\odot}, \label{eq:nsurv}
\end{eqnarray}
where, $M_{\rm vir}$ is the halo virial mass at $z=0$ 
with the overdensity according to \citet{Bryan1998}.

\begin{figure}
\centering 
\includegraphics[width=9.2cm]{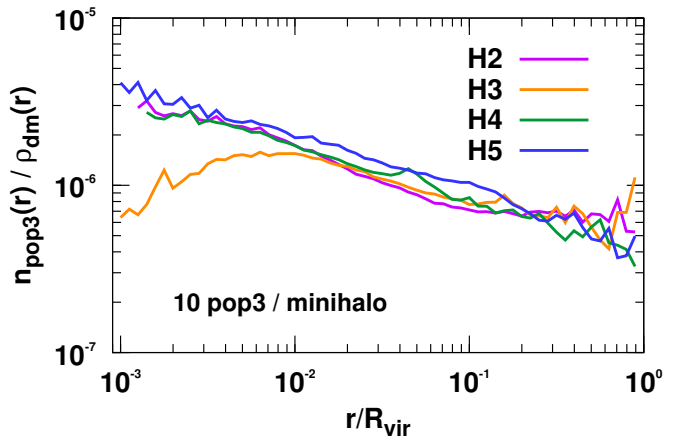}
\caption{ 
Ratio between radial number density profiles of \p\ survivors 
$n_{\rm surv}(r)$ and dark matter mass densities of host halos 
$\rho_{\rm dm}(r)$ for four Milky Way sized halos, H2 to H5. 
The smallest radii plotted 
are the reliability limits using criterions
suggested by \citet{Fukushige2001} and \citet{Power2003}.
}
\label{fig:dist_dm}
\end{figure}

\begin{table}[t]
\centering
\caption
{The number of particles $N$,  and the virial mass $M_{\rm vir}$ 
within the virial radius of four Milky Way sized halos in the simulation.
 }\label{tab1}
\begin{tabular}{lcc}
\hline\hline
Name  & $N$ & $M_{\rm vir} (10^{12} M_{\odot})$\\
\hline
H2 & 381,963,719 & 2.88 \\
H3 & 318,640,498 & 2.40 \\
H4 & 306,717,590 & 2.31 \\
H5 & 146,585,900 & 1.11 \\
\hline 
\end{tabular}
\end{table}

\begin{figure*}
\centering 
\includegraphics[width=6.3cm, angle=-90]{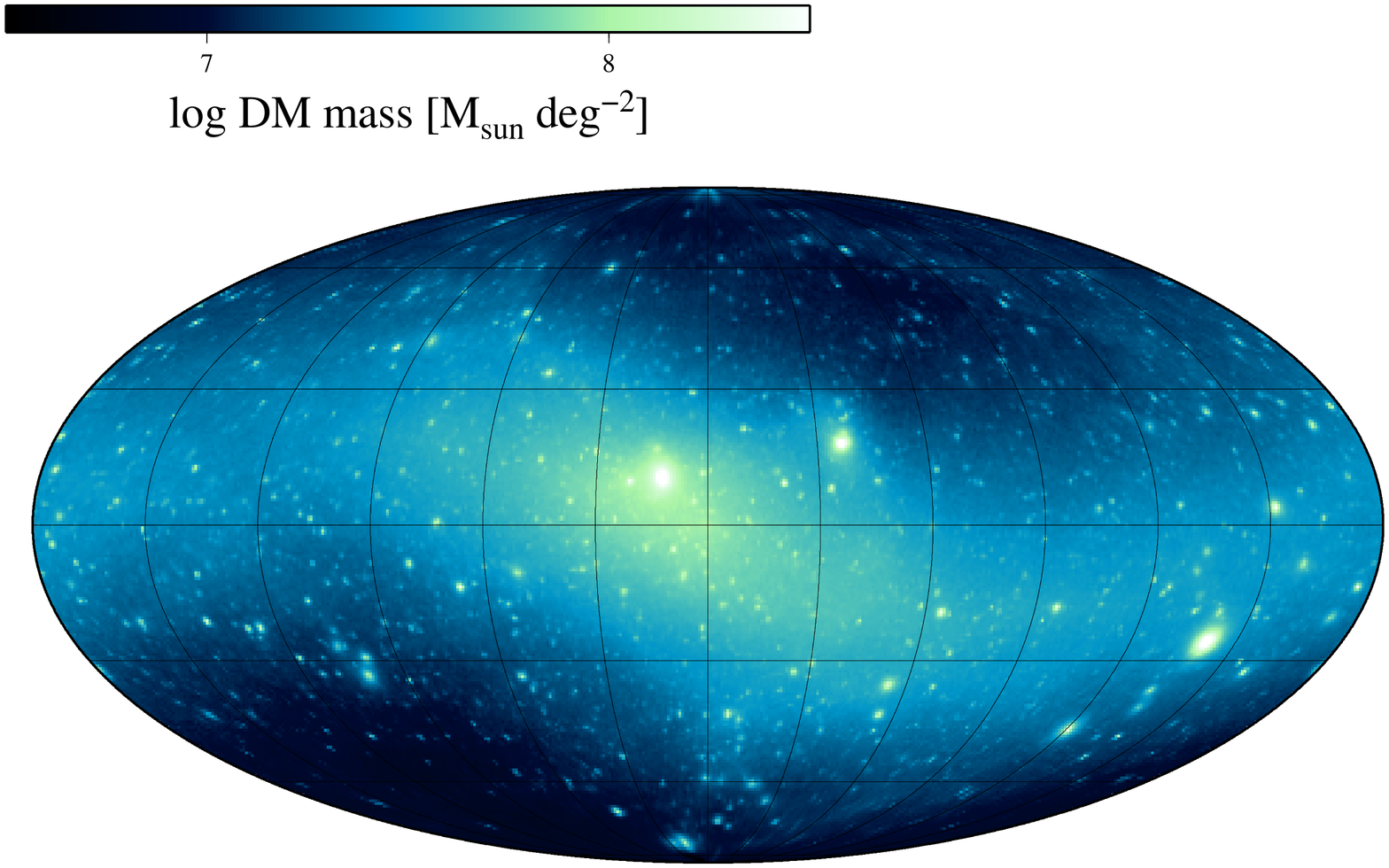} 
\includegraphics[width=6.3cm, angle=-90]{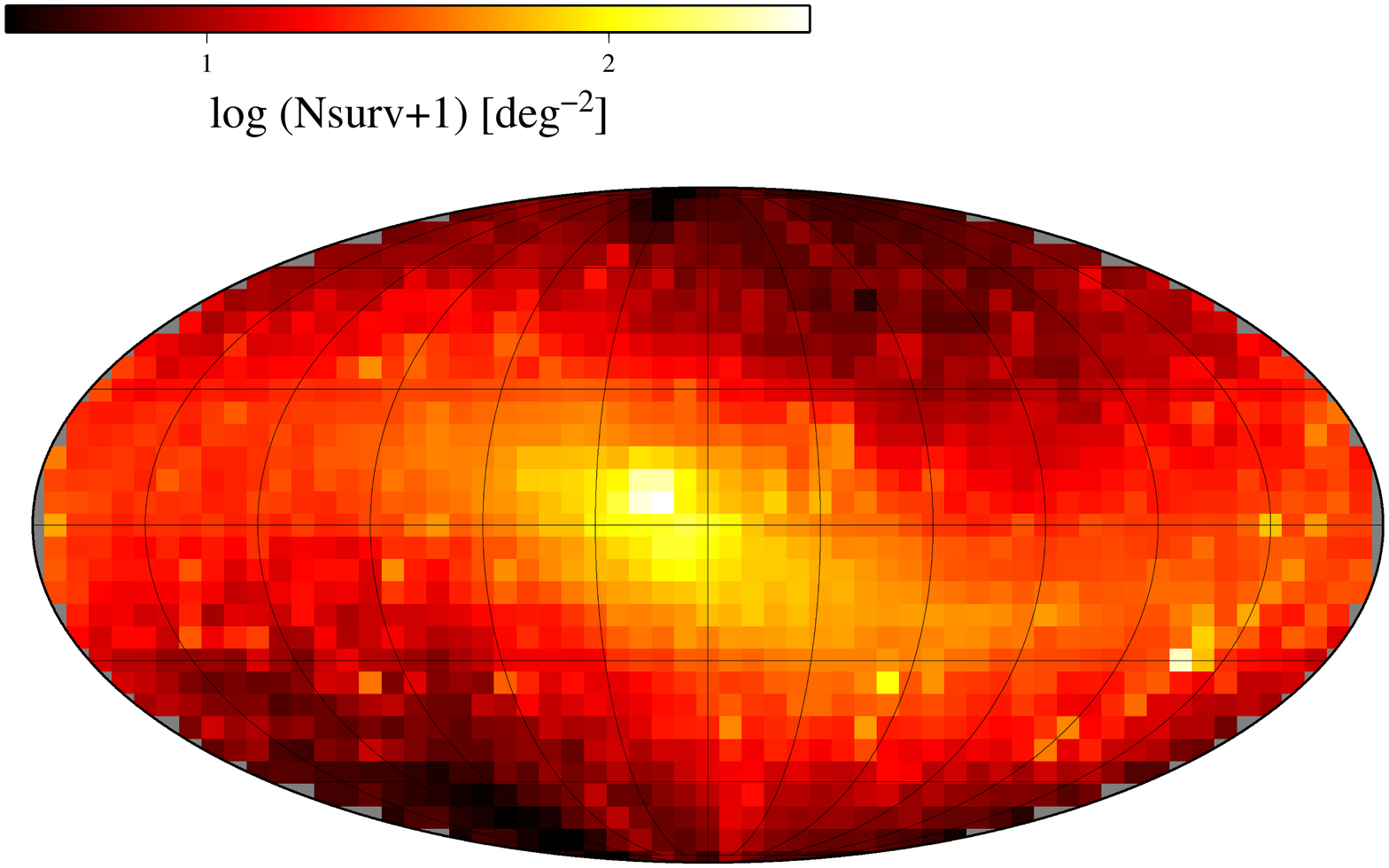} 
\includegraphics[width=6.3cm, angle=-90]{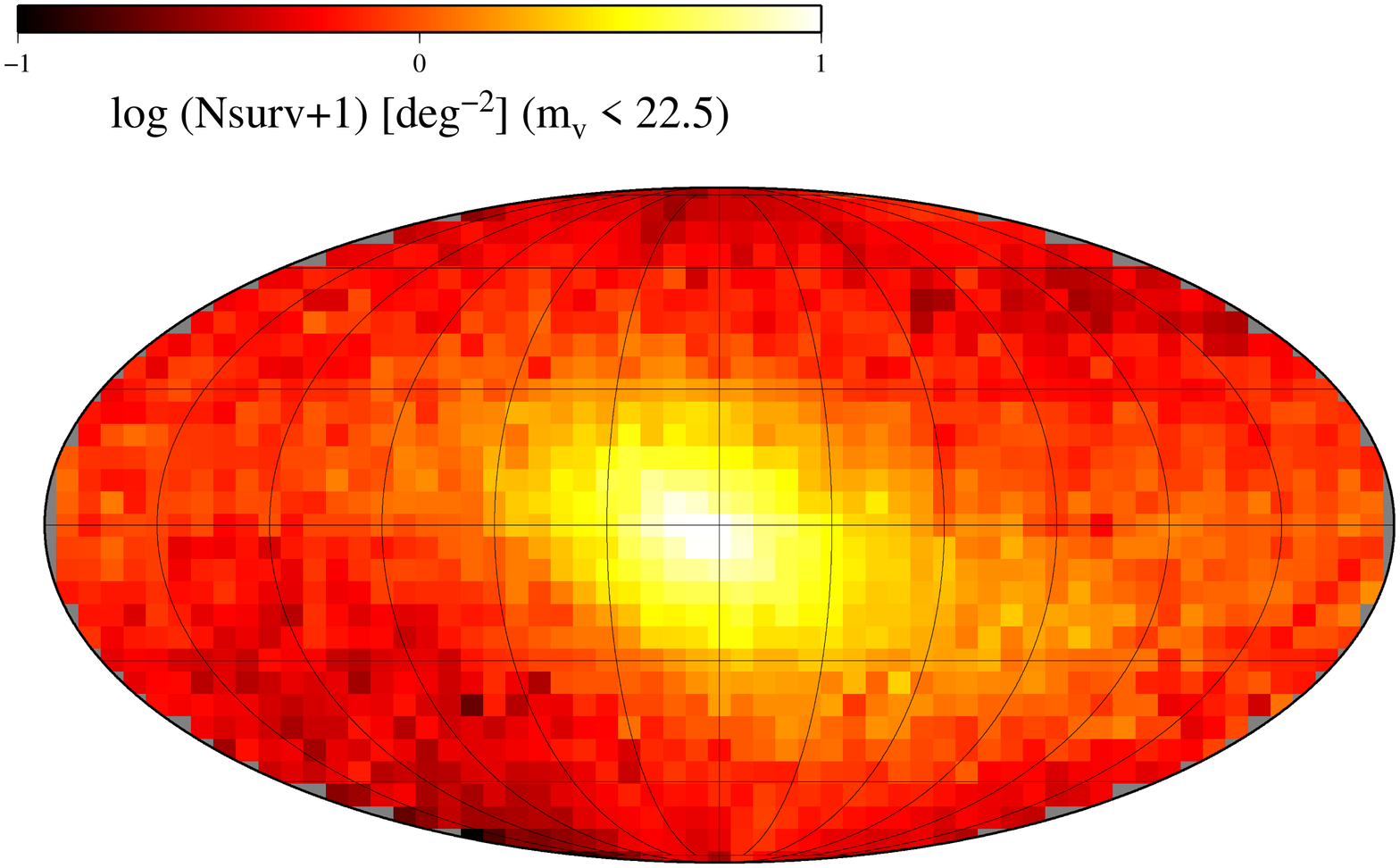} 
\includegraphics[width=6.3cm, angle=-90]{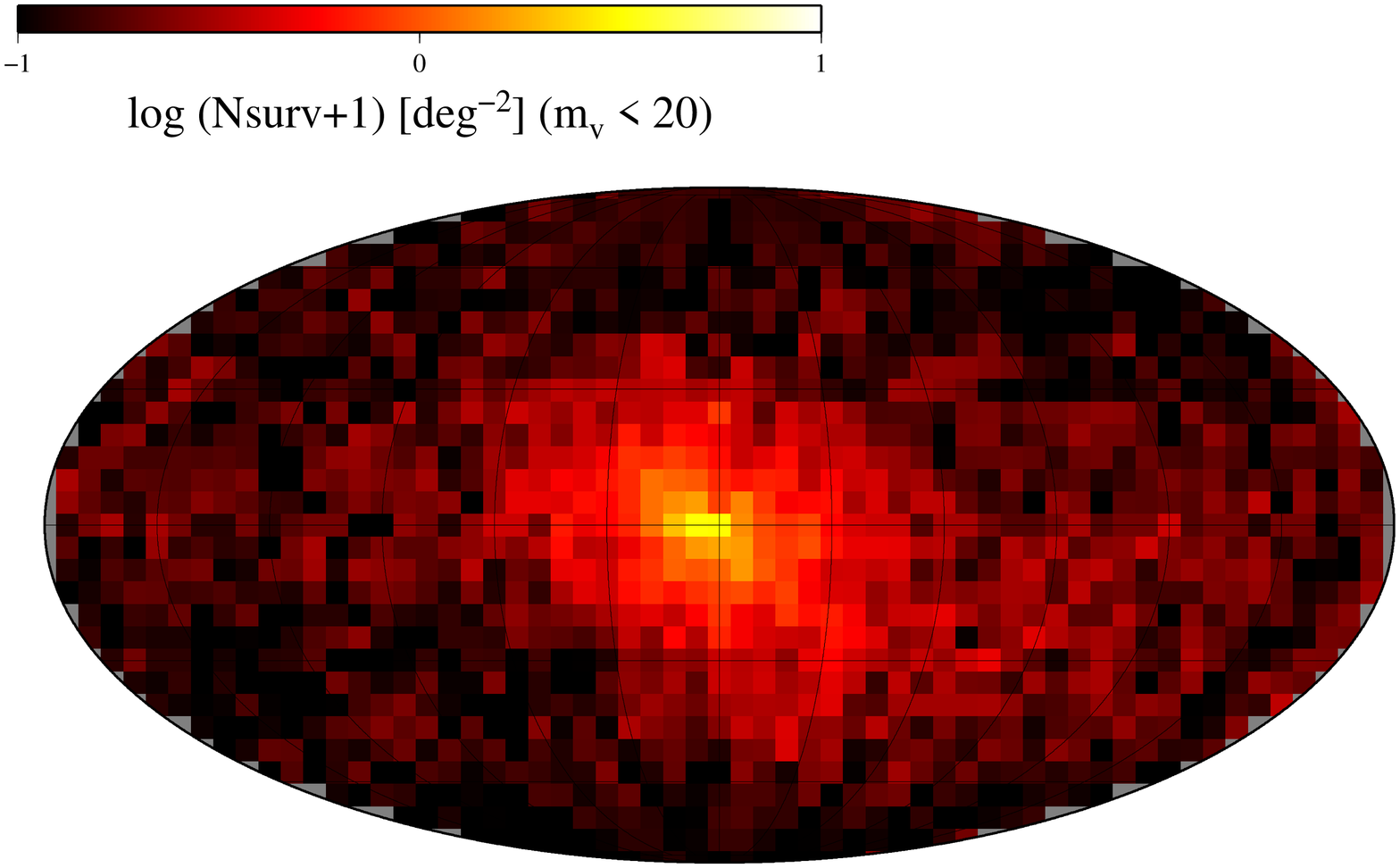} 
\caption{ 
All-sky map of the projected dark matter density and the number density 
of \p\ survivors in a Milky Way sized halo, H5. 
The top left panel is the dark matter density. 
The top right panel shows all survivors. 
The bottom left and bottom right panels show 
the number density of survivors 
with V-band magnitude brighter than 22.5 and 20.0, respectively.
}
\label{fig:map}
\end{figure*}

This fitting function indicates that the fixed number of survivors,
$10^{-7} \npopp$, per one solar dark matter mass exist in each halo
regardless of the halo mass, at least from $\sim 10^8 \Msun$
to the Milky Way mass.  This
dependence may be because the dependence of the halo formation history
(e.g., merger rate, mass accretion rate) on the halo mass is weak
\citep[e.g.,][]{Ishiyama2015}.

Hereafter, we focus on the spatial distribution of \p\ survivors 
in the Milky Way sized halos to discuss their observability, 
using the $\npopp=10$ model. 
In our simulation, four Milky Way sized halos are identified at $z=0$.
In table \ref{tab1}, the number of particles and the virial mass of
these halos are summarized.  These are second- to fifthmost-
massive halos in the entire simulation box.  Because the most massive halo (H1:
$5.09 \times 10^{12} \Msun$) is about twice more massive than the
Milky Way halo, we exclude this halo in the following analysis.

As described in \S \ref{sec:surv}, the spatial positions of survivors
in these halos are assumed to be those of tracers of dark matter
particles directly taken from the high resolution cosmological 
$N$-body simulation. 
This is of great advantage to early studies \citep[e.g.,][]{Komiya2015,Hartwig2015}, 
based on semi-analytic methods with merger trees extracted from 
extended Press-Schechter theory \citep[e.g.,][]{Press1974,Lacey1993}.
Other early studies
based on $N$-body simulations also suffered from insufficient mass
resolution to capture the formation of small minihalos
\citep[e.g.,][]{Tumlinson2010}.

In Figure \ref{fig:dist_dm}, we plot the ratio between radial number
density profiles of \p\ survivors $n_{\rm surv}(r)$ and dark
matter mass densities of host halos $\rho_{\rm dm}(r)$ for four Milky
way sized halos, H2 to H5.  
The survivors tend to be distributed in a manner  more concentrated than the dark matter, 
in particular, in the central regions ($r/R_{\rm vir} < 0.1$), 
except for H3. 
This implies that the formation history of H3 might 
be largely different from the other three halos. 

We made the all-sky map of the number density of \p\ 
survivors in these four halos.  The observer is located at 8.5 kpc from
the center of the halo.
Figure \ref{fig:map} shows all-sky maps of the projected dark matter 
density and the number density of all 
survivors per square degree for
H5.  Clearly, the distribution of survivors reflects that of the dark
matter as shown in Figure \ref{fig:dist_dm}.  The survivors tend to be
concentrated in the center of the halo and the subhalos.

The distributions of survivors with V-band magnitude brighter than
22.5 and 20.0 are also shown in Figure \ref{fig:map} and are pretty
different from that of all survivors.  
Note that these magnitudes are close to the limiting magnitudes of 
deep spectroscopic observations like PFS \citep{Takada2014}.
The concentrated distribution
in subhalos are invisible.  This is simply because the subhalos are
too far from the observer.  Even if the survivors in the subhalos are
bright stars, their apparent magnitude would be faint.

\section{Observation Strategy of \p\ Stars and Current Constraints}

\begin{figure}
\centering 
\includegraphics[width=6.5cm, angle=-90]{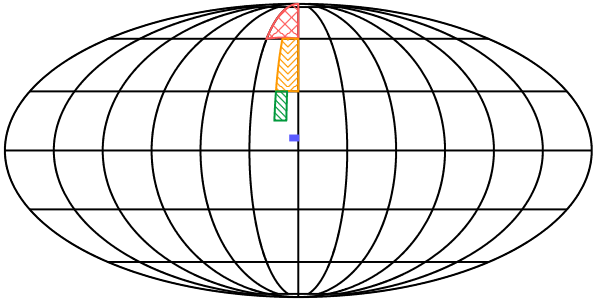}
\caption{
Four target fields in this study are highlighted as gray regions. 
From top to bottom, these regions correspond to 
high, middle, low latitude and central fields. 
}
\label{fig:target}
\end{figure}

\begin{figure*}
\centering 
\includegraphics[width=8.9cm]{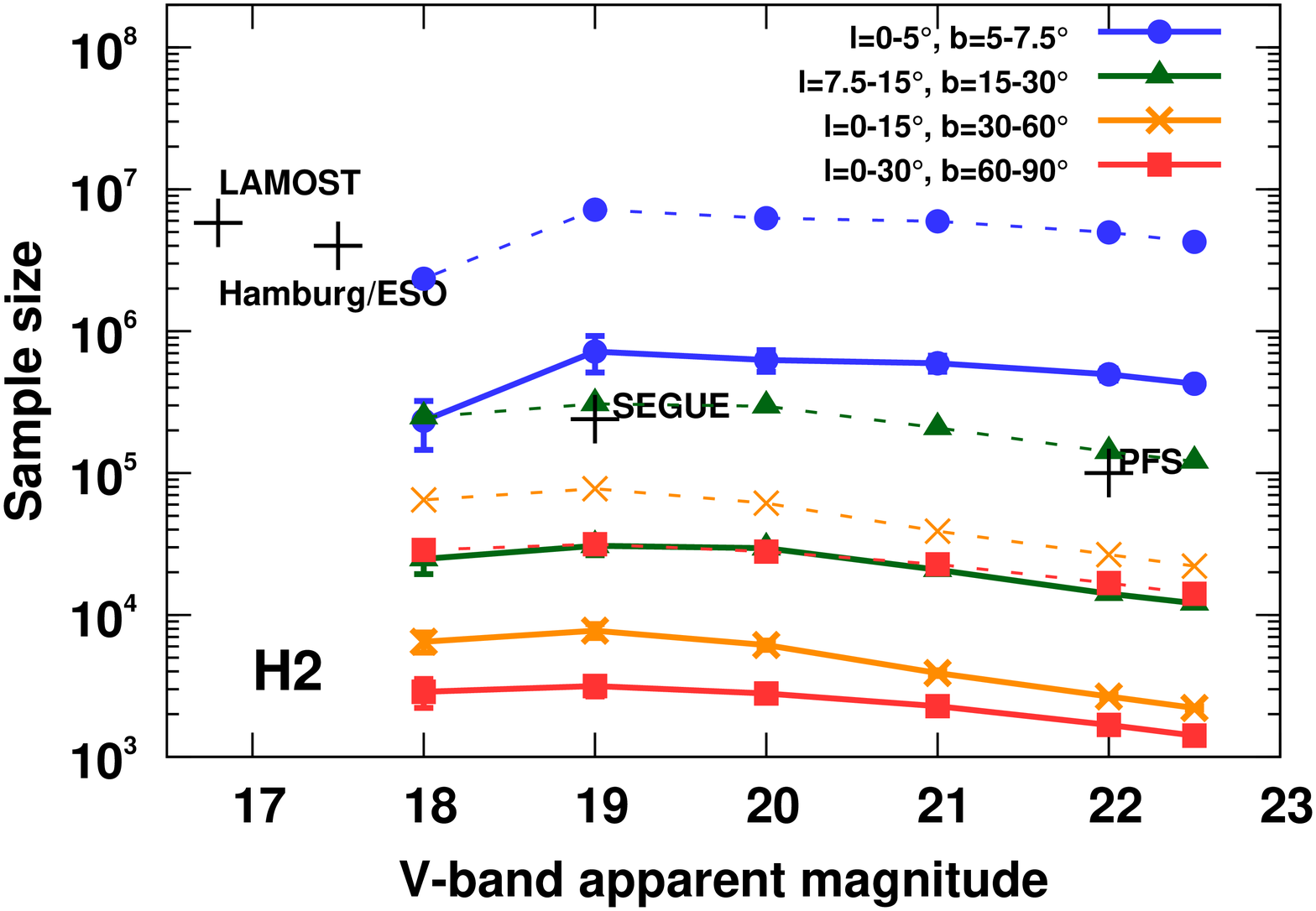} 
\includegraphics[width=8.9cm]{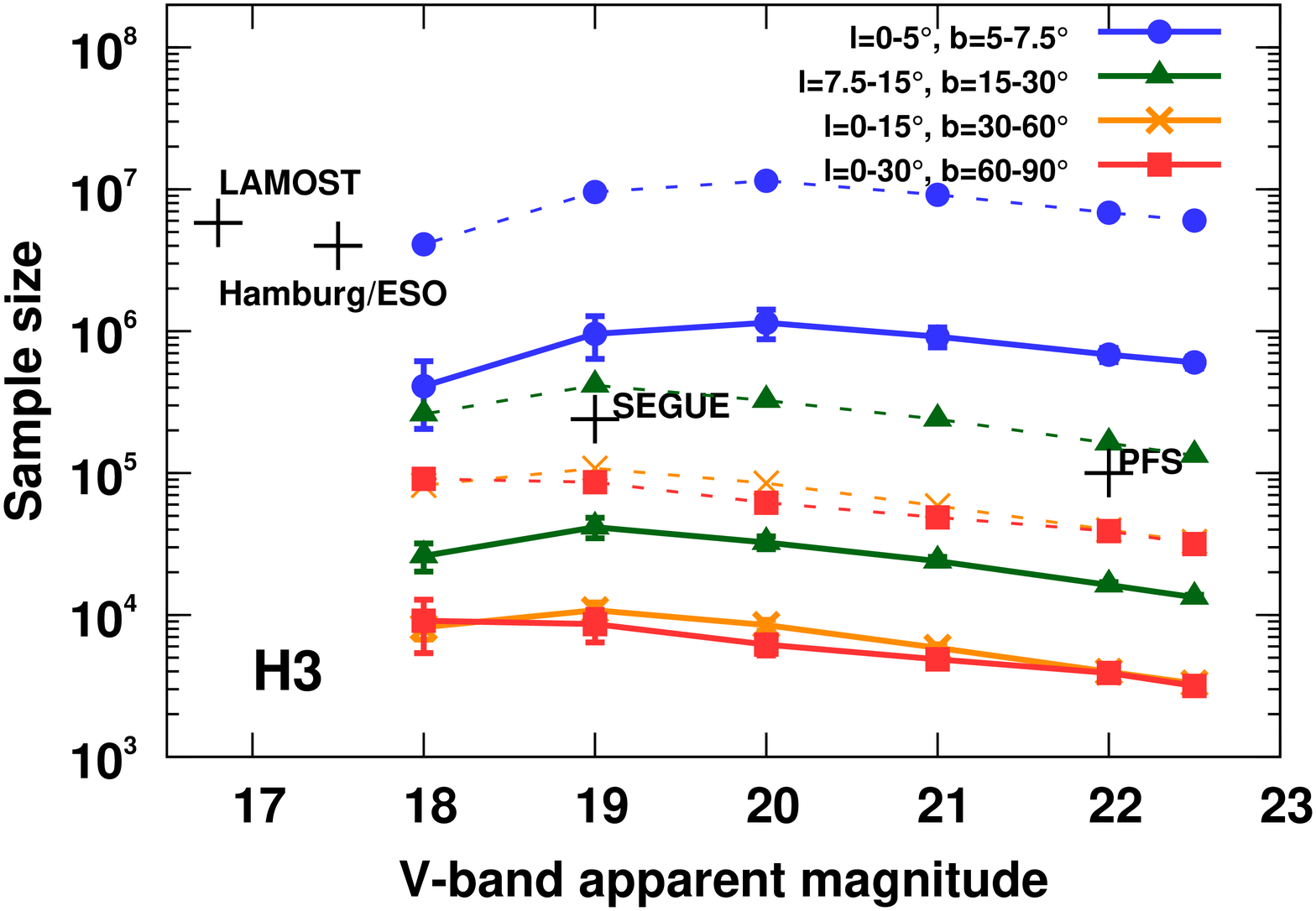} 
\includegraphics[width=8.9cm]{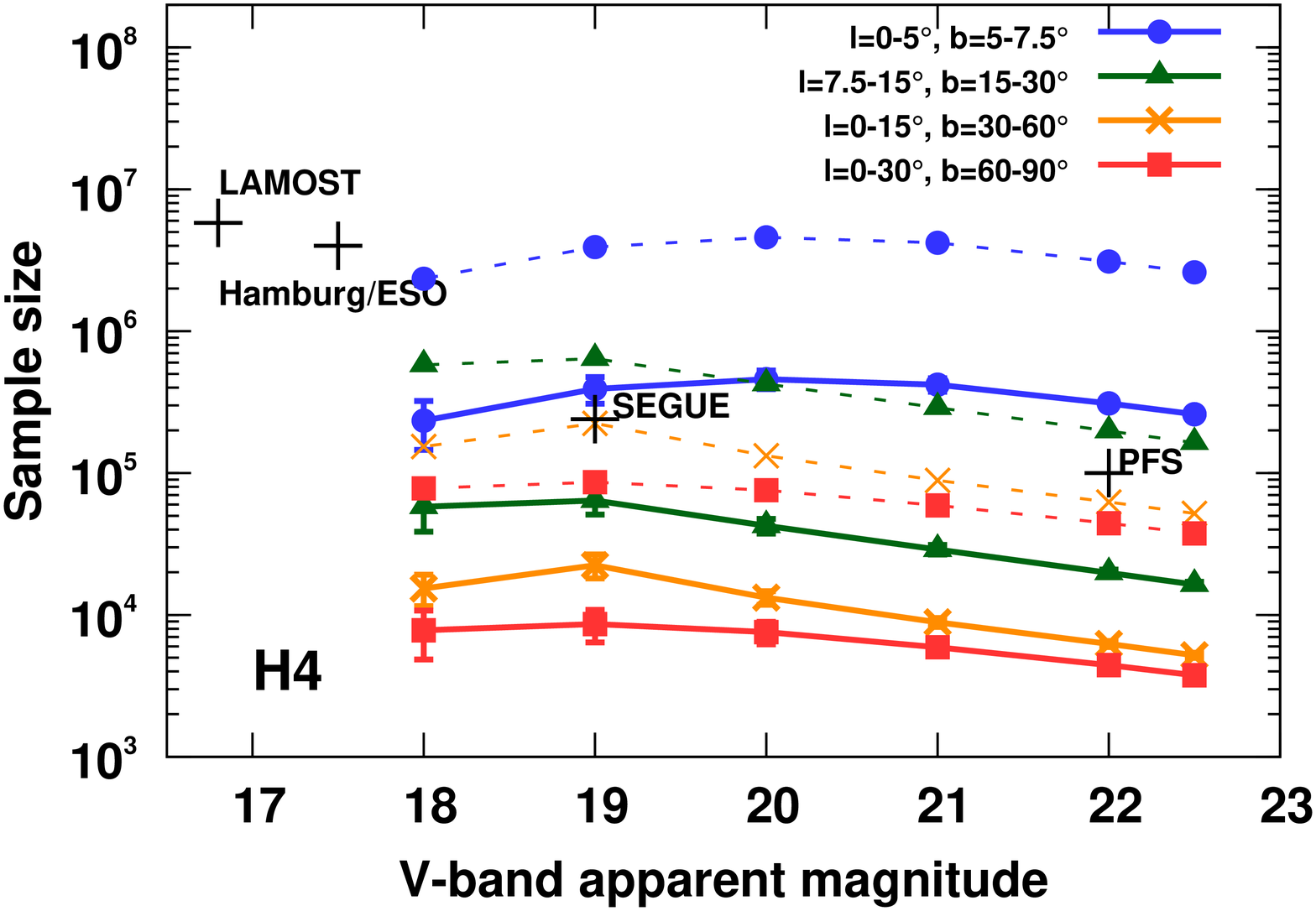} 
\includegraphics[width=8.9cm]{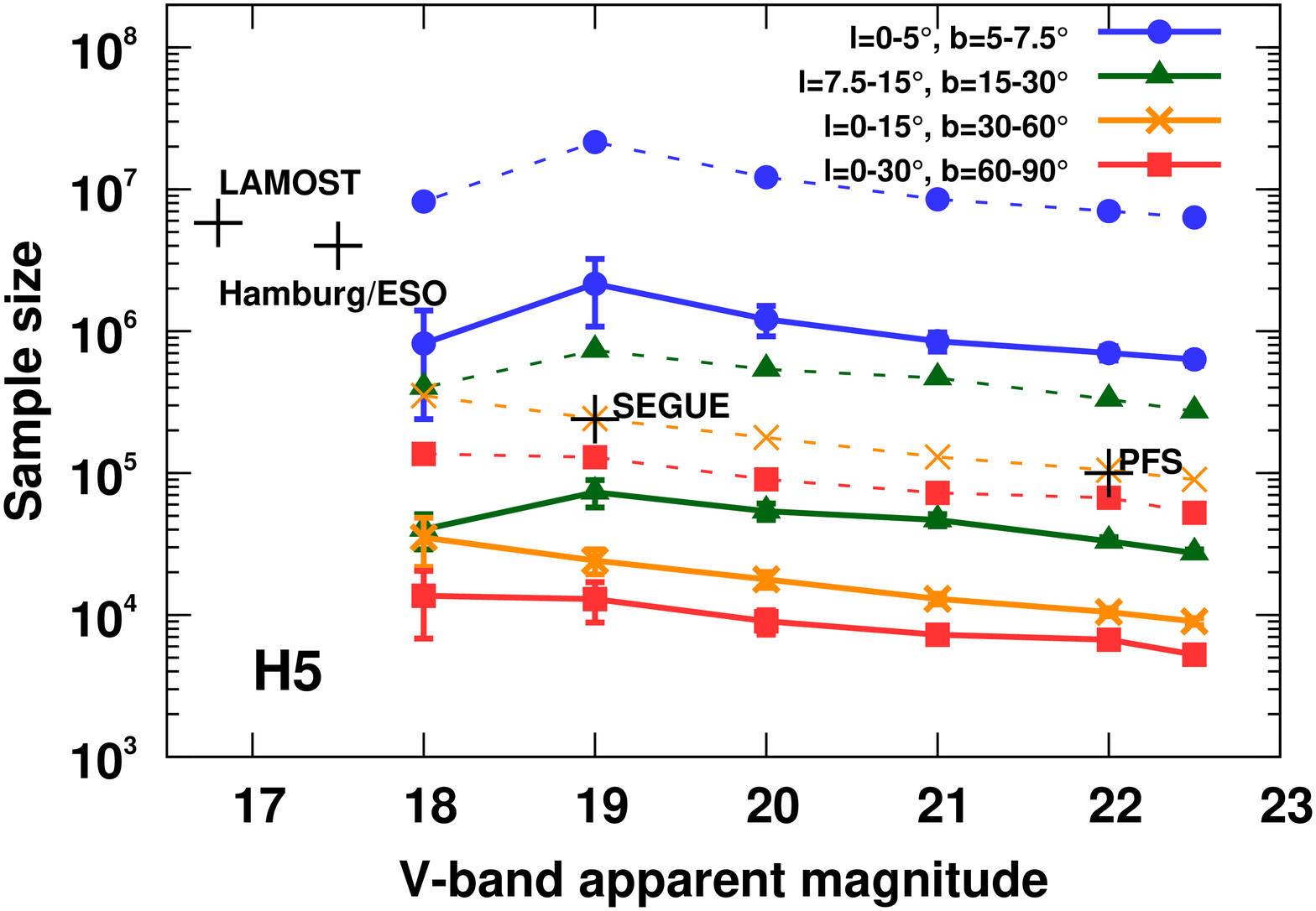} 
\caption{ 
Sample size to find one \p\ survivor in four Milky Way sized halos
as a function of V-band magnitude 
for four different target fields.
Solid curves show the results of the $\npopp=10$ model. 
Dashed curves are the results of this model multiplied 
by a factor of ten, which mimics the $\npopp=1$ model.
Crosses show observations, PFS \citep{Takada2014}, 
SEGUE \citep{Yanny2009}, 
Hamburg/ESO survey \citep{Christlieb2008},
and LAMOST \citep{Luo2015}, respectively. 
The error bars show their Poisson error.
Since most of survivors have a V-band magnitude, $m_{\rm V}>17$, 
we do not count field stars brighter than this value. 
}
\label{fig:obs}
\end{figure*}

\begin{figure}
\centering 
\includegraphics[width=9cm]{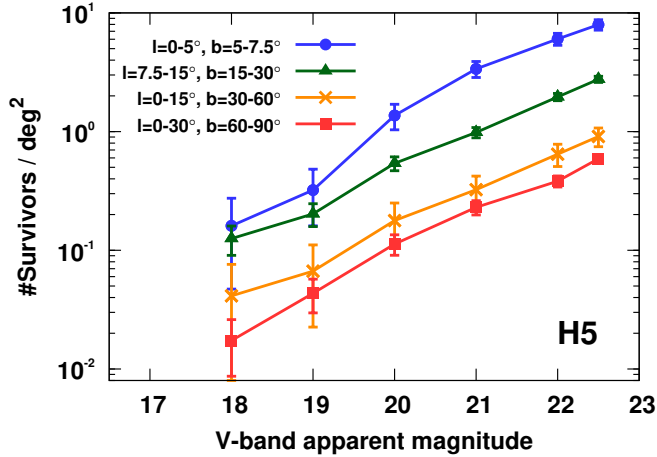}
\caption{ 
Number density of survivors brighter than the given V-band apparent magnitude 
per square degree in H5
for four different target fields.
The error bars show their Poisson error.
}
\label{fig:mv-dens}
\end{figure}

\begin{figure}
\centering 
\includegraphics[width=9cm]{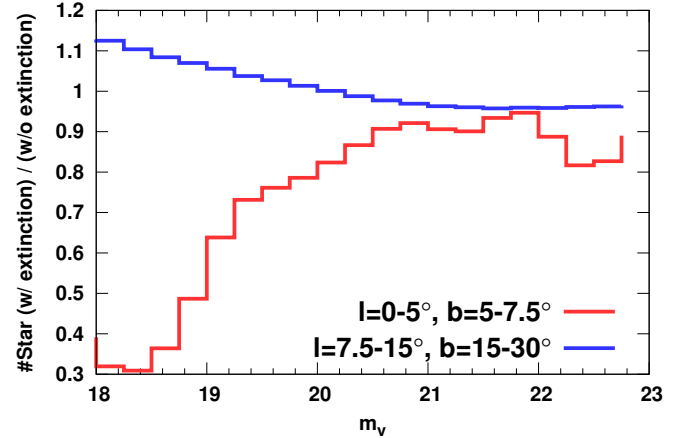}
\caption{ 
Effect of the ISM extinction on the number of field stars: 
the ratio between the number with and without the extinction 
in each V-band magnitude bin. 
}
\label{fig:extinction}
\end{figure}

\begin{figure*}
\centering 
\includegraphics[width=8.9cm]{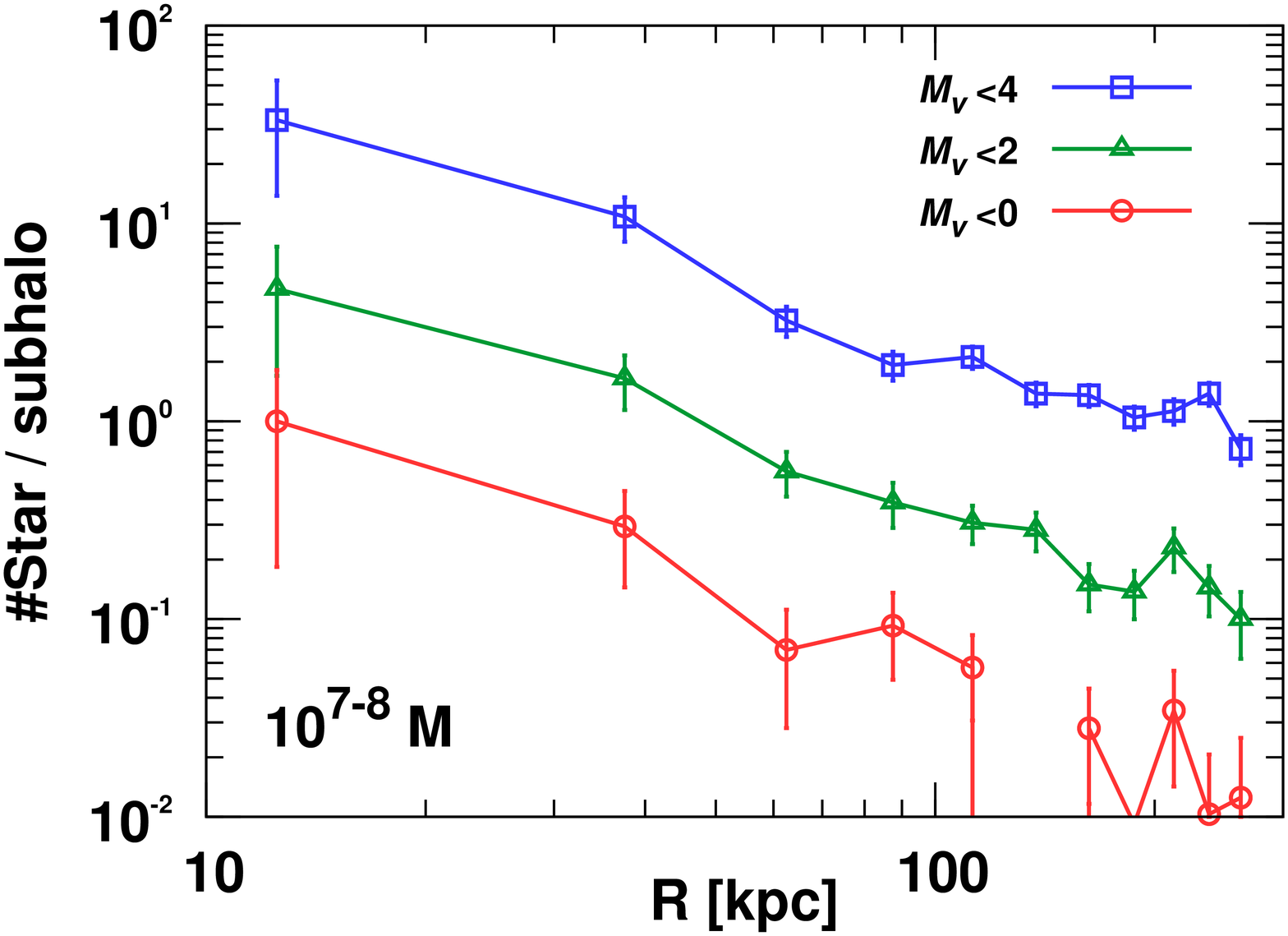}
\includegraphics[width=8.9cm]{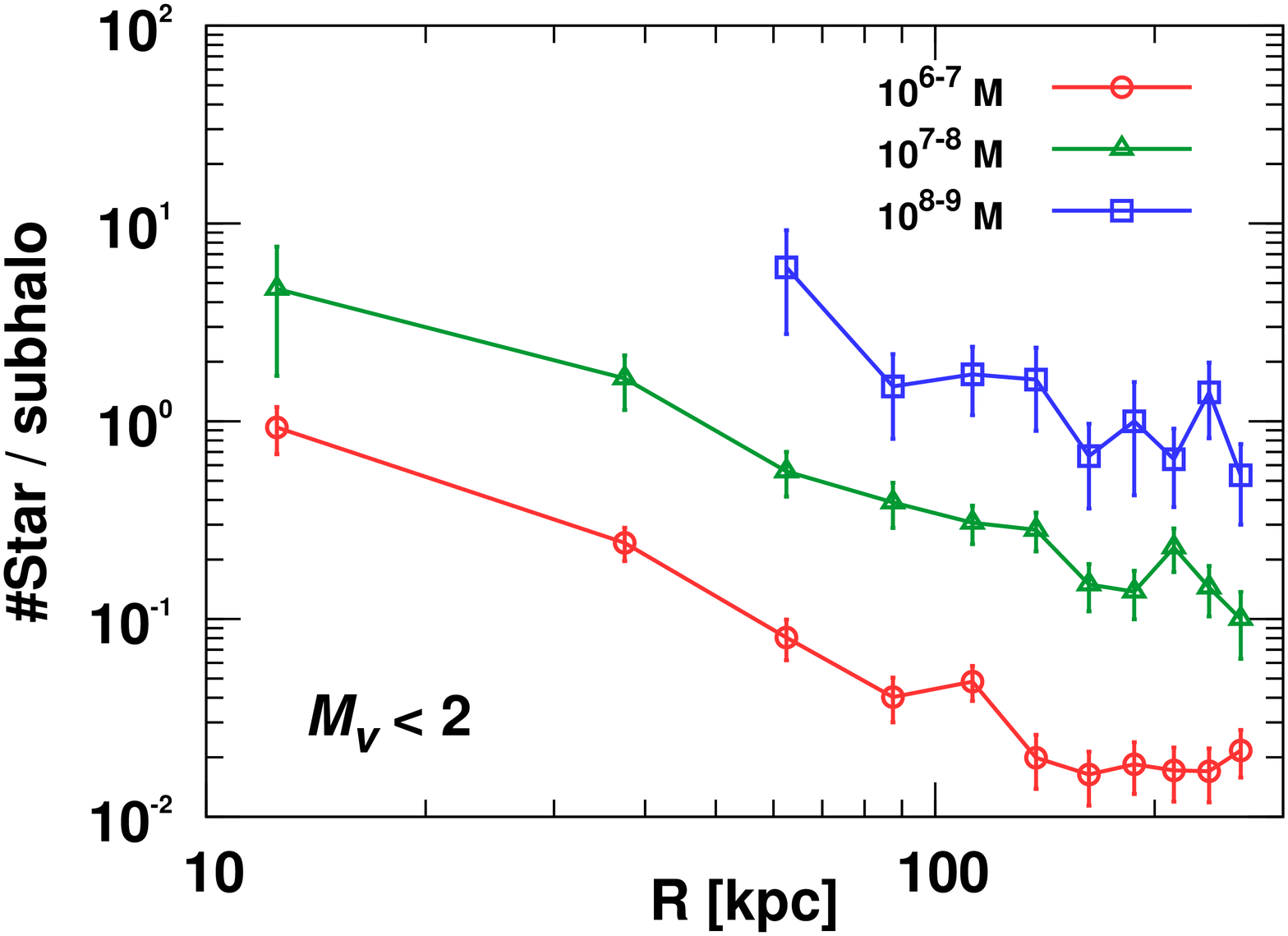}
\caption{ 
(Left)  
Radial profile of the average number of \p\ survivors 
brighter than the given absolute magnitudes in a subhalo of H5
with the mass range of $10^{7\mbox{\scriptsize --}8} M_\odot$ 
as a function of the radius from the halo center. 
(Right) Radial profile of the average number of \p\ survivors 
brighter than absolute magnitudes $M_{\rm V} = 2$ in a subhalo 
as a function of the radius from the halo center. 
Three curves correspond to three subhalo mass ranges. 
The error bars show their Poisson error.
 }
\label{fig:sub}
\end{figure*}

While the number of survivors becomes larger toward the center, the
stellar components of the Milky Way also show centrally concentrated
distribution. Thus, it is not trivial where the observability of
survivors is the largest. Previous studies, e.g., \cite{Hartwig2015},
had discussed this issue in galactic halo and bulge. However, they had
adopted a simple model to distinguish halo and bulge stars because
they used a semi-analytic model based on merger trees extracted from
extended Press-Schechter theory \citep[e.g.,][]{Press1974,Lacey1993},
which cannot predict the spatial distribution of halos. In contrast
to the previous studies, the combination of the cosmological
simulation and a \p\ star formation model provides the spatial
distribution of survivors in the Milky Way and satellite dwarf
galaxies. This enables us to discuss observation strategies to detect
\p\ survivors without an additional model.

\subsection{Milky Way}\label{sec:4.1}

The \p\ survivors in the Milky Way mix with other stellar
components of the Milky Way. Thus, the field stars need to be excluded
for the detection of the \p\ survivors. In order to evaluate
their detectability in different fields, we extract the stellar distribution and metallicity
distribution of the Milky Way using the online interface of the stellar
population synthesis model for the Milky Way, 
Besan\c{c}on\footnote{{\url http://model.obs-besancon.fr/}} \citep{Robin2003}.
The effect of the ISM extinction in the disk was not taken
into account. 

We calculate the number of survivors and field stars and the sample size
required to find one \p\ survivor in the following four
different fields:
\begin{enumerate}
\item high latitude field : 
$l=0^\circ$--$30^\circ$, $b=60^\circ$--$90^\circ$;
\item middle latitude field : 
$l=0^\circ$--$15^\circ$, $b=30^\circ$--$60^\circ$;
\item low latitude field : 
$l=7.5^\circ$--$15^\circ$, $b=15^\circ$--$30^\circ$; and
\item central field : 
$l=0^\circ$--$5^\circ$, $b=5^\circ$--$7.5^\circ$,
\end{enumerate}
which are displayed in Figure \ref{fig:target}.
These four fields represent the galactic halo, disk, bulge, and
center of the Milky Way.
The required sample size is defined as
the ratio of the number of field stars to that of survivors.
Since most of survivors have a V-band magnitude, $m_{\rm V}>17$, 
we do not count field stars brighter than this value. 

Figure \ref{fig:obs} shows the required sample size in four Milky Way sized halos as
a function of V-band limiting magnitude for four different target
fields. The result of the $\npopp=10$ model is shown
(solid curves). Dashed curves are the results of this model
multiplied by a factor of ten, which mimics the $\npopp=1$
model. The dependence of the number of survivors on
\npop\ shown in equation (\ref{eq:nsurv}) justifies this simple scaling.

In spite of the larger number of survivors toward the galactic center
(Figure \ref{fig:map}), higher latitude fields require lower sample sizes. This is
simply because the field stars are more concentrated at the low latitude
field than the survivors. The higher number density of field stars
at lower latitude makes the detection of the \p\ survivors
less efficient.
In the low latitude and central fields, 
the sample size required is
by a factor of $3\sim 5$ and $30\sim 200$
 larger than those in middle and
higher latitude fields, respectively. For all halos, the high latitude
field with the direction to the galactic plane 
($l=0^\circ$--$30^\circ$, $b=60^\circ$--$90^\circ$) are most efficient.

The required sample size is slightly reduced with deep limiting
magnitude. This is because the deep observation reaches the stars in the
galactic halo.
The fraction of survivors in the galactic halo is higher than the galactic disk
since the disk star shows more centrally concentrated distribution 
than that of dark matter.
Furthermore, the number of survivors per square degree is increasing
as we observe deeper, as indicated in Figure \ref{fig:mv-dens}, 
which gives the number density of survivors
brighter than the given V-band apparent magnitude per square degree in H5
for four different target fields.  However, the weak dependence on the
limiting magnitude indicates that a wide and shallow survey detects
more survivors than a deep and narrow survey with a given survey
power, i.e., telescope and instrument, as long as the number of
targets is sufficient.

When we take into account the ISM extinction, 
the expected number of both survivors and field stars 
becomes smaller, in particular, for lower latitude fields. 
We also extract the stellar distribution of the Milky Way 
with the extinction using the Besan\c{c}on 
\citep{Robin2003}~\footnote{
We set the extinction to 0.70 mag/kpc in the galactic disk.
}.
Figure \ref{fig:extinction} shows 
the effect of the ISM extinction on the number of field stars 
for the central and the low latitude fields, respectively;
there is little effect on the low latitude field
($l=7.5^\circ$--$15^\circ$, $b=15^\circ$--$30^\circ$).
On the other hand, 
the number of stars at the bright end is largely reduced 
for the central field ($l=0^\circ$--$5^\circ$, $b=5^\circ$--$7.5^\circ$).
Assuming that the extinction acts on survivors in the same manner
as the field stars, 
the number density of bright survivors ($m_{\rm V} < 19$) 
is reduced by a factor of $\sim$2 in only central field.

The observer position does not change the total number of survivors in
Milky Way sized halos but can influence those in these fields. To
investigate the effect, we calculate the number of survivors seen from
randomly placed 1,000 observers 
on a spherical surface lying 8.5 kpc from the halo center, 
for high and low latitudes fields
of H5, $l=0^\circ\range30^\circ$, $b = 60^\circ\range90^\circ$ and $l = 7.5^\circ\range15^\circ$,
$b = 15^\circ\range30^\circ$. 
In table \ref{tab2}, we summarize the median, 5, 25, 75, and 95
percentiles of the sample size for these fields. Regardless of the
V-band limiting magnitude $m_V$, the sample size of the high latitude
field ($l = 0^\circ\range30^\circ$, $b = 60^\circ\range90^\circ$) spread by a factor of nearly
two and four from 25 to 75 percentiles and from 5 to 95
percentiles. Those of the low latitude field ($l = 7.5^\circ\range15^\circ$,
$b = 15^\circ\range30^\circ$) spread by a factor of nearly 1.4 and two from 25 to
75 percentiles and from 5 to 95 percentiles. The 95 percentiles values
in the high latitude field are always larger than the 5 percentiles
values in the low latitude field.

The spectroscopic metal-poor surveys are mainly conducted at high
latitude fields so far. The results of cosmological simulations can be
directly compared with outcomes of the past and ongoing/planning
surveys. Figure \ref{fig:obs} also shows the sample size and limiting magnitude of
the following surveys with crosses; Hamburg/ESO survey
\citep{Christlieb2008}
, SEGUE
\citep{Yanny2009}, and future prospects by PFS \citep{Takada2014} and LAMOST
\citep{Li2015}. If we assume that Hamburg/ESO survey and SEGUE do not
detect any survivors and take into account the incompleteness 
of follow-up moderate dispersion spectroscopy for candidates
found by Hamburg/ESO survey \citep{Christlieb2006}, 
the formation model of \p\ stars with $\npopp=10$ is already excluded by these
observations
because the main targets of SEGUE are in 
relatively higher
latitude fields (Figure 1 in \cite{Yanny2009}).

There is a study that suggests that the surface pollution can
enhance the surface metal abundance of \p\ survivors up to
[Fe/H]~$\sim-5$ \citep{Komiya2015}, below which the Hamburg/ESO survey had
found two stars \citep{Christlieb2002,Frebel2005}. Even if their claim is
real, the outcomes of the past surveys still rule out the model with
$\npopp=10$ and favor $\npopp=1$.

\begin{table*}[t]
\centering
\caption{
Uncertainty of the sample size by various observational positions. 
The median, 5, 25, 75, and 95 percentiles of the 
sample size from 1,000 randomly placed observers are shown. 
$m_{\rm v}$ is the V-band magnitude cutoff. 
}
\label{tab2}
\begin{tabular}{lcccccc}
\hline\hline
Area & $m_{\rm v}$ & median & 5\% & 25\% & 75\% & 95\%\\
\hline
$l=0$--$30^\circ$, $b=60$--$90^\circ$ & 22 & $4.9 \times 10^3$  & $2.3 \times 10^3$ & $3.3 \times 10^3$ & $6.5 \times 10^3$ & $8.8 \times 10^3$ \\
$l=0$--$30^\circ$, $b=60$--$90^\circ$ & 20 & $7.8 \times 10^3$  & $4.4 \times 10^3$ & $5.7 \times 10^3$ & $1.1 \times 10^4$ & $1.6 \times 10^4$ \\
$l=7.5$--$15^\circ$, $b=15$--$30^\circ$ & 22 & $3.3 \times 10^4$  & $2.2 \times 10^4$ & $2.7 \times 10^4$ & $3.9 \times 10^4$ & $4.8 \times 10^4$ \\
$l=7.5$--$15^\circ$, $b=15$--$30^\circ$ & 20 & $6.0 \times 10^4$  & $4.0 \times 10^4$ & $5.1 \times 10^4$ & $7.4 \times 10^4$ & $9.4 \times 10^4$ \\
\hline 
\end{tabular}
\end{table*}

Recently, Skymapper performs a photometric metal-poor survey with a
narrow-band filter with a bandpass corresponding to \ion{Ca}{2} HK
lines and broad-band filters \citep{Keller2007}. Skymapper
photometrically excludes the metal-rich field stars and successfully
found the most iron-deficient star \citep{Keller2014} and an extremely
metal-poor star in the bulge \citep{Howes2015}. Although the efficiency
and completeness of the photometric classification have not been
presented, the stars with [Fe/H]~$>-1.5$ is largely excluded (Figure~2
in \citealt{Howes2014}).  Assuming that the photometric classification
could exclude $95\%$ of stars with [Fe/H]~$>-1.5$ and adopting the
metallicity distribution of the Milky Way model, the numbers of field
stars in the high, middle, low latitude, and central fields are reduced
by factors of $2.8$, $3.8$, $7.8$, and $17$, respectively. The
photometric classification enhances the efficiency of follow-up
spectroscopy by these factors. As a result, the required sample sizes
are comparable in the high and middle latitude fields, while the
required sample sizes in the low latitude and central fields are still
2 times and 20 times larger than that of the high latitude field,
respectively. 
Practically, the most efficient field and depth depend
on the number of fibers and the field of view of instruments for
spectroscopy. If the photometric classification works well, while the
high latitude field is most efficient for instruments with low fiber
density, e.g., HERMES ($\sim120$~fibers/deg$^2$, \citealt{Sheinis2014})
for GALAH survey \citep[e.g.,][]{DeSilva2015} and LAMOST
($\sim800$~fibers/deg$^2$, \citealt{Li2015}), the middle latitude field
is most efficient for instruments with high fiber density, e.g., PFS
($\sim1800$~fibers/deg$^2$, \citealt{Takada2014}).  By performing spectroscopic
observation of a million of stars after the photometric classification
in the future, we can constrain the low mass \p\ star IMF with
unprecedented accuracy.

\subsection{Dwarf Galaxies}

Only bright survivors are reachable in dwarf galaxies but the
distribution of survivors is concentrated (Figure~\ref{fig:map}). Thus, the
detection of survivors in dwarf galaxies could be more efficient than
that in the Milky Way. In this subsection, we investigate the
possibility of the detection of \p\ survivors in dwarf
galaxies. However, the known missing satellite problem 
\citep[e.g.,][]{Klypin1999, Moore1999, Ishiyama2009}
deters the direct prediction from the cosmological simulations and thus we
evaluate the average number of \p\ survivors 
in the model with $\npopp=10$, 
brighter than
given absolute magnitudes in a subhalo (left panel of Figure \ref{fig:sub}). 
The thresholds correspond to absolute magnitudes of the tip of redgiant stars
($M_V=0$), the redgiant stars ($M_V=2$), and the turn-off stars
($M_V=4$). 
We identify subhalos in each halo with the
ROCKSTAR phase space halo/subhalo finder \citep{Behroozi2013}.

While the average number is larger for fainter thresholds, dependence of
the average number on the distance from the galactic center is similar.
The average number decreases with the distance because the distant
subhalos were recently formed and do not satisfy the criteria of \p\
star formation (Sec.~2.2). Furthermore, the average number per subhalo is
larger for more massive subhalos
(right panel of Figure \ref{fig:sub})
because the more massive subhalos
include more minihalos with \p\ star formation. 
The inverse of the average number per subhalo is the required number of
dwarf galaxies to find one \p\ survivor,
if subhalos hosting dwarf galaxies do not preferentially
contain a larger number of survivors than all other subhalos. 
The required number of dwarf galaxies is estimated in less than ten 
at $<100$~kpc for the tip of redgiant stars 
(corresponding to $m_V\sim20$), 
if the dwarf galaxies have a common mass of $\sim10^7 M_\odot$ 
\citep[e.g.,][]{Strigari2008, Okamoto2009} 
although there is a debate \citep{Hayashi2012}.

The number of dwarf galaxies discovered so far is 
nearly 30 \citep[e.g.,][]{Koposov2015, Bechtol2015}.
Although the extremely metal-poor stars with [Fe/H]~$<-3$ have
been discovered especially in ultra faint dwarf galaxies
\citep[e.g.,][]{Frebel2014}, no \p\ survivor has been
detected. This constraint disfavors $n_{\rm pop3}=10$ but is consistent
with $n_{\rm pop3}=1$.

Recently, the wide-field photometric surveys newly found dwarf galaxies
including ultra faint dwarf galaxies at $<100$~kpc
\citep[e.g.,][]{Belokurov2007,Laevens2015}. In addition to the large
average number of \p\ survivors in nearby dwarf galaxies, the faint threshold can be
realized for nearby dwarf galaxies. Thus, the all sky survey of nearby
dwarf galaxies is highly demanded for the detection of \p\
survivors. The increasing number of nearby dwarf galaxies refines the
constraint on $n_{\rm pop3}$.

In contrast to the \p\ survivors in the Milky Way, the
survivors in dwarf galaxies are as faint as $m_V\sim20$ and the numbers
of the detectable \p\ survivors and dwarf galaxies strongly
depend on the limiting magnitude. Therefore, the light collecting power
is essential on the detection of survivors in dwarf galaxies. Although
the number of stars that are bright enough for the high-dispersion
spectroscopy with $8$m class telescopes is small, candidates of the
survivors are reached with the narrow-band imaging and/or
low/medium-dispersion spectroscopy with $8$-m class telescopes, and will
be good targets for the follow-up high-dispersion spectroscopy with
$30$m class telescopes.

\section{Discussion}

\subsection{Comparison with Other Studies}

\citet{Hartwig2015} provided
the expected number of \p\ survivors in the Milky Way.
They reported that an unbiased survey of $4\times 10^6$  halo stars could
impose a constraint on the low mass end of the \p\ IMF. 
This value is more than two orders of magnitude larger 
than our $\npopp=10$ model (Figure \ref{fig:obs}). 
Even in the $\npopp=1$ model, more than an order of magnitude difference
exists. 
They used a simple model to distinguish halo and bulge stars
because they adopted a semi-analytic model 
based on merger trees extracted from extended Press-Schechter theory, 
which could not give the spatial distribution of halos.

In the model of \citet{Hartwig2015}, 
there are physical processes we do not include in our model, 
dynamical heating of gas during mass accretion and mergers. 
However, they reported that the main mechanisms
to suppress \p\ star formation are 
metal enrichment at low redshifts and the threshold 
mass of the minihalo at high redshifts, 
both of which are included in our model by different manners. 

The main reason for such large difference is the adoption of the IMF of \p stars.
Whereas \citet{Hartwig2015} used a logarithmically flat IMF from 
0.01 to 100$\Msun$, we adopted the Kroupa IMF from 0.15 to 1.0 $\Msun$ for
low mass \p stars.  Consequently, the number of low mass \p\ stars per
minihalo is also largely different.  In their model, 
the average number was 
about 0.1 for an IMF in the mass range 0.65 to 100 $\Msun$
(T. Hartwig 2015, private communication), 
which is nearly two orders of magnitude smaller than
our fixed $\npopp=10$ model.  This large difference is sufficient to
explain large gaps in the estimation of the sample sizes needed to
constrain on the low mass end of the IMF. 
From Figure \ref{fig:obs}, 
it is easy to infer that 
the large difference in the estimation of sample sizes would disappear
when we used the $\npopp=0.1$ model, consistent with \citet{Hartwig2015}. 
This means that our and their results
qualitatively agree with each other 
although there is quantitative disagreement.

Why is the average number of \p\ stars per minihalo 
in both models so different?
The decisive difference is that the model of 
\citet{Hartwig2015} used the 
lowest mass of the IMF as an arbitrary parameter and
tried to constrain it whereas 
the number of \p\ stars is used in our model. 
In fact, the lower bound of the IMF 0.65$\Msun$ in their model 
was obtained to match the sample size of Hamburg/ESO survey.

The radial number density of survivors (Figure
\ref{fig:dist_dm}) distributes more concentrated than the dark matter and 
qualitatively agrees with \citet{Gao2010} (see 
also \citet{White2000, Diemand2005b, Tumlinson2010}), 
although the concentration of our results are less than 
that of \citet{Gao2010}. 
This quantitative disagreement can be 
explained by the difference of cosmological parameters. 
The parameter $\sigma_8$ of \citet{Gao2010} is 0.9, 
while 0.83 is used in our simulation. 
The Press-Schechter theory predicts 
the number of halos of $M \sim 10^6\Msun$ at $z=25$ in $\sigma_8=0.9$
by a factor of $\sim3$ larger than $\sigma_8=0.83$. 
Because earlier formed progenitor halos tend to concentrate on the center of halos, 
larger $\sigma_8$ should result in a higher concentration of survivors.

One may imagine that the difference can be 
explained by the way to select tracers of the low mass \p\ stars.
Whereas we use randomly selected dark matter particles
in minihalos, the most bound
particles were used in \citet{Gao2010}.  
Since \p\ stars should be born in the central dense regions of minihalos,
they might stay there with a high probability.  However, in such
situations, \p\ stars could not be long-lived low mass stars because of
subsequent mass accretion.  On the other hand, some of stars could be
kicked away from the central regions of minihalos to regions with
shallower potential, via the gravitational many body interaction
\citep[e.g., ][]{Clark2011, Clark2011b, Smith2011, Greif2011,
  Greif2012, Umemura2012, Susa2013, Machida2013}, 
As a consequence, they could be
long-lived low mass stars because of the poor mass accretion, and
easily stripped from minihalos by tides of larger halos.  Our method
to select \p\ tracers mimics this latter process.

To investigate how the distribution of survivors 
is sensitive to the choice of tracers,
we selected the most bound particle from each minihalo 
as tracers and calculated the radial number 
density distributions of survivors. 
Compared with the randomly selected model ($\npopp=1$), 
we confirm that both distributions agree well.
This suggests that  the criteria
of \p\ forming minihalos 
and adopted cosmological parameters
are more important for the distributions of survivors, 
and cosmological radiation hydrodynamical simulations with 
large volume are necessary to calibrate models.

\subsection{Other Physical Processes}

In this study, we neglect some physical processes that 
can affect the number of survivors. 
We will take them into account for our model in future. 

Surfaces of \p\ survivors could be polluted with metals
\citep[e.g.,][]{Yoshii1981, Shigeyama2003, Frebel2009, Komiya2010,
  Komiya2015}.  \citet{Frebel2009} studied the accretion of 
the ISM on a star in the galactic disk and
concluded that it is generally negligible.  On the other hand,
\citet{Komiya2015} investigated the effect of the ISM accretion in
minihalos by a semi-analytic model based on the hierarchical
clustering scenario.  They demonstrated that the pollution is
effective in the early stage of the hierarchical formation of halos
and the surface iron abundance of survivors could be enhanced up to 
$\rm [Fe/H] \sim -5$.  The Fe abundance of hyper metal poor stars (HMP) can be
explained by this pollution scenario.  
Even if we take into account this pollution effect, the models with
$\npopp=10$ are still ruled out because only two stars of [Fe/H]$<-5$
have been discovered so far. Thus, the main constraint of the \p\
star formation model in our study is not much affected by this effect
as discussed in \S \ref{sec:4.1}.

The abundance of the minihalos could be suppressed by the streaming
velocities, that is, the supersonic relative velocity of baryon and
dark matter arises at the time of recombination
\citep{Tseliakhovich2010, Tseliakhovich2011}.  Recent numerical
simulations including the streaming velocity indicate that typical
minihalo mass increases by a factor of three and \p\ star formation is
delayed by $\Delta z \sim 4$ \citep{Greif2011b}.  On the other hand,
\citet{Stacy2011} suggest that there is little effect on the gas
evolution by the typical streaming velocity.  Currently the effect of
the streaming velocity is highly uncertain, and thus, it is needed to
evaluate it accurately through large \p\ formation simulations.

\subsection{Remnants of Massive \p\ stars}
In the present paper, we have discussed on the observational
possibility of finding \p\ survivors. The effort to search these stars
will help to constrain the low mass end of the \p\ IMF 
at $M \la 1M_\odot$ severely.
On the other hand, the theoretical IMF of \p\ stars extends to $\sim
1000M_\odot$ \citep{Susa2014,Hirano2014,Hirano2015}.
Thus, it is worth mentioning the observations to be compared 
with the high mass part of the \p\ IMF.


Metal poor stars could have been born in the remnants of \p\
stars. Hence the comparison of the abundance ratios in the atmospheres
of these stars with the theoretical predictions of the nucleosynthesis
in the \p\ stars could have great significance in constraining the
high mass part of the IMF. In fact, the observed abundance ratios
provide little evidence of pair instability supernovae (PISNe), which
should be found if some \p\ stars form in the range of $140M_\odot
\la M \la 260M_\odot$. It is rather consistent with the assumption
that most of the \p\ stars are less massive than $100M_\odot$ to
supply the metals by core collapse supernovae \citep{Susa2014}. We have
to keep in mind that we cannot directly conclude the less massive
($\la 100M_\odot$) IMF is favored, because the lack of 
PISNe pattern
is only evident in very metal poor stars with [Fe/H] $\leqq$ -3, and
has not been proved for [Fe/H]$>-3$. 
In fact, at [Fe/H]=$-2.5$ \citet{Aoki2014} found a
possible candidate of a second generation star formed in a
remnant of a massive star with $\ga100\Msun$.
Further observations of the stars with higher
metallicity will give us more information on \p\ IMF.
Present high resolution simulations will
be coupled with semi-analytical models of low metallicity star formation
to be compared with these observations in the near future.


Some theoretical calculations on the \p\ star formation naturally predict
formation of multiple stellar systems including massive binaries 
\citep{Stacy2012,Susa2013,Stacy2014,Susa2014}. Such systems will evolve into
black hole binaries, which would merge to form a more massive black
hole by emitting the gravitational waves. \citet{Kinugawa2015} estimated
the detection rate of such events to be pretty high($\sim$180
$\rm events~yr^{-1}$). If the gravitational wave from such objects are
detected at a predicted rate, 
it will be a circumstantial
evidence that we are witnessing the merging of \p\ black holes.
In fact, the recent discovery of the gravitational wave from
$29M_\odot-36M_\odot$ black hole binary coalescence \citep{Abbott2016}
suggests a high rate of such events. The forthcoming data release will
provide a better estimate of the frequency.
Meanwhile, the high resolution cosmological simulations as
presented in this paper will be coupled with \p\ binary formation/evolution
theory to give a better prediction of the event rate from the theoretical side.

\section{Summary}

If low mass \p\ stars are less massive than 0.8$M_\odot$, their 
lifetime is longer than the cosmic time, and thus they could survive to
be found in the Milky Way.
We have studied the number and the distribution of low mass
\p\ survivors in the Milky Way by combining a large cosmological $N$-body
simulation and a \p\ formation model.  Unlike early studies, we can
predict the spatial distribution of survivors in the Milky Way by
simulating both hierarchical formation of dark matter minihalos and
Milky Way halos.  We model the \p\ formation in H$_2$ cooling
minihalos without metal under UV radiation of the Lyman-Werner bands.
Assuming a Kroupa IMF from 0.15 to 1.0 $\Msun$ for low mass \p\, as a
working hypothesis, we try to constrain the theoretical models in
reverse through current and future observations.

From the mass and the collapse redshift of \p\ stars, 
we calculated the magnitude of various bands using an isochrone model.
We selected randomly \npop\ dark matter particles from each minihalo
as tracers of the low mass \p\ stars. 
We used $\npopp = 1$ and 10 models.
The spatial positions of the tracers at $z=0$ 
are assumed to be those of \p\ survivors. 

We find that 
the survivors tend to
concentrate on the center of halo and subhalos.
 We also derived the sample size
  required to find one \p\ survivor and compare it with past
  metal-poor star surveys.  Since the number density of stars in the
  galactic disk is too large to negate the increase of the number of
  survivors toward the galactic center, higher latitude fields require
  lower sample sizes to detect survivors.  If we assume that available
  observations have not detected any survivors, the formation model of
  low mass \p\ stars with more than ten stars per minihalo is already
  excluded. 

We also consider practical observation strategies of \p\
survivors in the Milky Way and the dwarf galaxies. 
The photometric classification with optimized 
narrow band can largely enhance the efficiency.
Provided that the photometric classification
could exclude $95\%$ of stars with [Fe/H]~$>-1.5$, 
the numbers of field
stars in the high, middle, low, latitude and central fields are reduced
by factors of $2.8$, $3.8$, $7.8$, and $17$, respectively.
As a result, the required sample sizes
are comparable in the high and middle latitude fields, while the
required sample sizes in the low latitude and central fields are still
2 times and 20 times larger than that of the high latitude field,
respectively. 

The required number of
dwarf galaxies to find one \p\ survivor 
is estimated in less than ten 
at $<100$~kpc for the tip of redgiant stars 
(corresponding to $m_V\sim20$). 
Assuming no \p\ survivor has been
detected, not $n_{\rm pop3}=10$ but 
$n_{\rm pop3}=1$ is favored, consistent with 
the current observations of the Milky Way.
The all sky survey of nearby
dwarf galaxies are highly demanded for the detection of \p\
survivors and refines the constraint on the low mass \p\ IMF.

We also discuss the way to constrain the IMF of \p\ stars at a high mass range of 
$M \ga 10 \Msun$. 
For $M < 260 \Msun$, surveys for the metal poor stars at [Fe/H]$\ga-3$ could find 
the trace of PISNe 
abundance ratio, if the \p\ IMF extends to the mass range of PISNe as the theories predict.

\acknowledgements
We thank the anonymous referee for his/her valuable comments.
We thank Tilman Hartwig and Wako Aoki for fruitful discussions. 
Numerical computations were partially carried out on Aterui
supercomputer at Center for Computational Astrophysics, CfCA, of
National Astronomical Observatory of Japan, and the K computer at the
RIKEN Advanced Institute for Computational Science (Proposal numbers
hp140212 and hp150226).  This work has been funded by MEXT HPCI
STRATEGIC PROGRAM. 
We thank the support by MEXT/JSPS KAKENHI grant No. 15H01030 (TI) 
and 22540295 (HS).

\bibliographystyle{apj}

\begin{thebibliography}{}
\expandafter\ifx\csname natexlab\endcsname\relax\def\natexlab#1{#1}\fi

\bibitem[{{Abbott} {et~al.}(2016){Abbott}, {Abbott}, {Abbott}, {Abernathy},
  {Acernese}, {Ackley}, {Adams}, {Adams}, {Addesso}, {Adhikari}, \&
  et~al.}]{Abbott2016}
{Abbott}, B.~P., {Abbott}, R., {Abbott}, T.~D., {et~al.} 2016, Physical Review
  Letters, 116, 061102

\bibitem[{{Abel} {et~al.}(2002){Abel}, {Bryan}, \& {Norman}}]{Abel2002}
{Abel}, T., {Bryan}, G.~L., \& {Norman}, M.~L. 2002, Science, 295, 93

\bibitem[{{Ahn} {et~al.}(2012){Ahn}, {Iliev}, {Shapiro}, {Mellema}, {Koda}, \&
  {Mao}}]{Ahn2012}
{Ahn}, K., {Iliev}, I.~T., {Shapiro}, P.~R., {et~al.} 2012, \apjl, 756, L16

\bibitem[{{Aoki} {et~al.}(2014){Aoki}, {Tominaga}, {Beers}, {Honda}, \&
  {Lee}}]{Aoki2014}
{Aoki}, W., {Tominaga}, N., {Beers}, T.~C., {Honda}, S., \& {Lee}, Y.~S. 2014,
  Science, 345, 912

\bibitem[{{Bechtol} {et~al.}(2015){Bechtol}, {Drlica-Wagner}, {Balbinot},
  {Pieres}, {Simon}, {Yanny}, {Santiago}, {Wechsler}, {Frieman}, {Walker},
  {Williams}, {Rozo}, {Rykoff}, {Queiroz}, {Luque}, {Benoit-L{\'e}vy},
  {Tucker}, {Sevilla}, {Gruendl}, {da Costa}, {Fausti Neto}, {Maia}, {Abbott},
  {Allam}, {Armstrong}, {Bauer}, {Bernstein}, {Bernstein}, {Bertin}, {Brooks},
  {Buckley-Geer}, {Burke}, {Carnero Rosell}, {Castander}, {Covarrubias},
  {D'Andrea}, {DePoy}, {Desai}, {Diehl}, {Eifler}, {Estrada}, {Evrard},
  {Fernandez}, {Finley}, {Flaugher}, {Gaztanaga}, {Gerdes}, {Girardi},
  {Gladders}, {Gruen}, {Gutierrez}, {Hao}, {Honscheid}, {Jain}, {James},
  {Kent}, {Kron}, {Kuehn}, {Kuropatkin}, {Lahav}, {Li}, {Lin}, {Makler},
  {March}, {Marshall}, {Martini}, {Merritt}, {Miller}, {Miquel}, {Mohr},
  {Neilsen}, {Nichol}, {Nord}, {Ogando}, {Peoples}, {Petravick}, {Plazas},
  {Romer}, {Roodman}, {Sako}, {Sanchez}, {Scarpine}, {Schubnell}, {Smith},
  {Soares-Santos}, {Sobreira}, {Suchyta}, {Swanson}, {Tarle}, {Thaler},
  {Thomas}, {Wester}, {Zuntz}, \& {DES Collaboration}}]{Bechtol2015}
{Bechtol}, K., {Drlica-Wagner}, A., {Balbinot}, E., {et~al.} 2015, \apj, 807,
  50

\bibitem[{{Beers} \& {Christlieb}(2005)}]{Beers2005}
{Beers}, T.~C., \& {Christlieb}, N. 2005, \araa, 43, 531

\bibitem[{{Behroozi} {et~al.}(2013){Behroozi}, {Wechsler}, \&
  {Wu}}]{Behroozi2013}
{Behroozi}, P.~S., {Wechsler}, R.~H., \& {Wu}, H.-Y. 2013, \apj, 762, 109

\bibitem[{{Belokurov} {et~al.}(2007){Belokurov}, {Zucker}, {Evans}, {Kleyna},
  {Koposov}, {Hodgkin}, {Irwin}, {Gilmore}, {Wilkinson}, {Fellhauer},
  {Bramich}, {Hewett}, {Vidrih}, {De Jong}, {Smith}, {Rix}, {Bell}, {Wyse},
  {Newberg}, {Mayeur}, {Yanny}, {Rockosi}, {Gnedin}, {Schneider}, {Beers},
  {Barentine}, {Brewington}, {Brinkmann}, {Harvanek}, {Kleinman}, {Krzesinski},
  {Long}, {Nitta}, \& {Snedden}}]{Belokurov2007}
{Belokurov}, V., {Zucker}, D.~B., {Evans}, N.~W., {et~al.} 2007, \apj, 654, 897

\bibitem[{{Bromm} {et~al.}(2002){Bromm}, {Coppi}, \& {Larson}}]{Bromm2002}
{Bromm}, V., {Coppi}, P.~S., \& {Larson}, R.~B. 2002, \apj, 564, 23

\bibitem[{{Bromm} \& {Larson}(2004)}]{Bromm2004}
{Bromm}, V., \& {Larson}, R.~B. 2004, \araa, 42, 79

\bibitem[{{Bryan} \& {Norman}(1998)}]{Bryan1998}
{Bryan}, G.~L., \& {Norman}, M.~L. 1998, \apj, 495, 80

\bibitem[{{Caffau} {et~al.}(2011){Caffau}, {Bonifacio}, {Fran{\c c}ois},
  {Sbordone}, {Monaco}, {Spite}, {Spite}, {Ludwig}, {Cayrel}, {Zaggia},
  {Hammer}, {Randich}, {Molaro}, \& {Hill}}]{Caffau2011}
{Caffau}, E., {Bonifacio}, P., {Fran{\c c}ois}, P., {et~al.} 2011, \nat, 477,
  67

\bibitem[{{Christlieb}(2006)}]{Christlieb2006}
{Christlieb}, N. 2006, in Astronomical Society of the Pacific Conference
  Series, Vol. 353, Stellar Evolution at Low Metallicity: Mass Loss,
  Explosions, Cosmology, ed. H.~J.~G.~L.~M. {Lamers}, N.~{Langer}, T.~{Nugis},
  \& K.~{Annuk}, 271

\bibitem[{{Christlieb} {et~al.}(2008){Christlieb}, {Sch{\"o}rck}, {Frebel},
  {Beers}, {Wisotzki}, \& {Reimers}}]{Christlieb2008}
{Christlieb}, N., {Sch{\"o}rck}, T., {Frebel}, A., {et~al.} 2008, \aap, 484,
  721

\bibitem[{{Christlieb} {et~al.}(2002){Christlieb}, {Bessell}, {Beers},
  {Gustafsson}, {Korn}, {Barklem}, {Karlsson}, {Mizuno-Wiedner}, \&
  {Rossi}}]{Christlieb2002}
{Christlieb}, N., {Bessell}, M.~S., {Beers}, T.~C., {et~al.} 2002, \nat, 419,
  904

\bibitem[{{Clark} {et~al.}(2008){Clark}, {Glover}, \& {Klessen}}]{Clark2008}
{Clark}, P.~C., {Glover}, S.~C.~O., \& {Klessen}, R.~S. 2008, \apj, 672, 757

\bibitem[{{Clark} {et~al.}(2011{\natexlab{a}}){Clark}, {Glover}, {Klessen}, \&
  {Bromm}}]{Clark2011b}
{Clark}, P.~C., {Glover}, S.~C.~O., {Klessen}, R.~S., \& {Bromm}, V.
  2011{\natexlab{a}}, \apj, 727, 110

\bibitem[{{Clark} {et~al.}(2011{\natexlab{b}}){Clark}, {Glover}, {Smith},
  {Greif}, {Klessen}, \& {Bromm}}]{Clark2011}
{Clark}, P.~C., {Glover}, S.~C.~O., {Smith}, R.~J., {et~al.}
  2011{\natexlab{b}}, Science, 331, 1040

\bibitem[{{Crocce} {et~al.}(2006){Crocce}, {Pueblas}, \&
  {Scoccimarro}}]{Crocce2006}
{Crocce}, M., {Pueblas}, S., \& {Scoccimarro}, R. 2006, \mnras, 373, 369

\bibitem[{{Davis} {et~al.}(1985){Davis}, {Efstathiou}, {Frenk}, \&
  {White}}]{Davis1985}
{Davis}, M., {Efstathiou}, G., {Frenk}, C.~S., \& {White}, S.~D.~M. 1985, \apj,
  292, 371

\bibitem[{{De Silva} {et~al.}(2015){De Silva}, {Freeman}, {Bland-Hawthorn},
  {Martell}, {de Boer}, {Asplund}, {Keller}, {Sharma}, {Zucker}, {Zwitter},
  {Anguiano}, {Bacigalupo}, {Bayliss}, {Beavis}, {Bergemann}, {Campbell},
  {Cannon}, {Carollo}, {Casagrande}, {Casey}, {Da Costa}, {D'Orazi}, {Dotter},
  {Duong}, {Heger}, {Ireland}, {Kafle}, {Kos}, {Lattanzio}, {Lewis}, {Lin},
  {Lind}, {Munari}, {Nataf}, {O'Toole}, {Parker}, {Reid}, {Schlesinger},
  {Sheinis}, {Simpson}, {Stello}, {Ting}, {Traven}, {Watson}, {Wittenmyer},
  {Yong}, \& {{\v Z}erjal}}]{DeSilva2015}
{De Silva}, G.~M., {Freeman}, K.~C., {Bland-Hawthorn}, J., {et~al.} 2015,
  \mnras, 449, 2604

\bibitem[{{Diemand} {et~al.}(2005){Diemand}, {Madau}, \&
  {Moore}}]{Diemand2005b}
{Diemand}, J., {Madau}, P., \& {Moore}, B. 2005, \mnras, 364, 367

\bibitem[{{Frebel} {et~al.}(2009){Frebel}, {Johnson}, \& {Bromm}}]{Frebel2009}
{Frebel}, A., {Johnson}, J.~L., \& {Bromm}, V. 2009, \mnras, 392, L50

\bibitem[{{Frebel} {et~al.}(2014){Frebel}, {Simon}, \& {Kirby}}]{Frebel2014}
{Frebel}, A., {Simon}, J.~D., \& {Kirby}, E.~N. 2014, \apj, 786, 74

\bibitem[{{Frebel} {et~al.}(2005){Frebel}, {Aoki}, {Christlieb}, {Ando},
  {Asplund}, {Barklem}, {Beers}, {Eriksson}, {Fechner}, {Fujimoto}, {Honda},
  {Kajino}, {Minezaki}, {Nomoto}, {Norris}, {Ryan}, {Takada-Hidai},
  {Tsangarides}, \& {Yoshii}}]{Frebel2005}
{Frebel}, A., {Aoki}, W., {Christlieb}, N., {et~al.} 2005, \nat, 434, 871

\bibitem[{{Fukushige} \& {Makino}(2001)}]{Fukushige2001}
{Fukushige}, T., \& {Makino}, J. 2001, \apj, 557, 533

\bibitem[{{Fuller} \& {Couchman}(2000)}]{Fuller2000}
{Fuller}, T.~M., \& {Couchman}, H.~M.~P. 2000, \apj, 544, 6

\bibitem[{{Gao} {et~al.}(2010){Gao}, {Theuns}, {Frenk}, {Jenkins}, {Helly},
  {Navarro}, {Springel}, \& {White}}]{Gao2010}
{Gao}, L., {Theuns}, T., {Frenk}, C.~S., {et~al.} 2010, \mnras, 403, 1283

\bibitem[{{Girardi} {et~al.}(2000){Girardi}, {Bressan}, {Bertelli}, \&
  {Chiosi}}]{Girardi2000}
{Girardi}, L., {Bressan}, A., {Bertelli}, G., \& {Chiosi}, C. 2000, \aaps, 141,
  371

\bibitem[{{Greif} {et~al.}(2012){Greif}, {Bromm}, {Clark}, {Glover}, {Smith},
  {Klessen}, {Yoshida}, \& {Springel}}]{Greif2012}
{Greif}, T.~H., {Bromm}, V., {Clark}, P.~C., {et~al.} 2012, \mnras, 424, 399

\bibitem[{{Greif} {et~al.}(2011{\natexlab{a}}){Greif}, {Springel}, {White},
  {Glover}, {Clark}, {Smith}, {Klessen}, \& {Bromm}}]{Greif2011}
{Greif}, T.~H., {Springel}, V., {White}, S.~D.~M., {et~al.} 2011{\natexlab{a}},
  \apj, 737, 75

\bibitem[{{Greif} {et~al.}(2011{\natexlab{b}}){Greif}, {White}, {Klessen}, \&
  {Springel}}]{Greif2011b}
{Greif}, T.~H., {White}, S.~D.~M., {Klessen}, R.~S., \& {Springel}, V.
  2011{\natexlab{b}}, \apj, 736, 147

\bibitem[{{Haiman} {et~al.}(1996){Haiman}, {Thoul}, \& {Loeb}}]{Haiman1996}
{Haiman}, Z., {Thoul}, A.~A., \& {Loeb}, A. 1996, \apj, 464, 523

\bibitem[{{Hartwig} {et~al.}(2015){Hartwig}, {Bromm}, {Klessen}, \&
  {Glover}}]{Hartwig2015}
{Hartwig}, T., {Bromm}, V., {Klessen}, R.~S., \& {Glover}, S.~C.~O. 2015,
  \mnras, 447, 3892

\bibitem[{{Hayashi} \& {Chiba}(2012)}]{Hayashi2012}
{Hayashi}, K., \& {Chiba}, M. 2012, \apj, 755, 145

\bibitem[{{Hirano} {et~al.}(2015){Hirano}, {Hosokawa}, {Yoshida}, {Omukai}, \&
  {Yorke}}]{Hirano2015}
{Hirano}, S., {Hosokawa}, T., {Yoshida}, N., {Omukai}, K., \& {Yorke}, H.~W.
  2015, \mnras, 448, 568

\bibitem[{{Hirano} {et~al.}(2014){Hirano}, {Hosokawa}, {Yoshida}, {Umeda},
  {Omukai}, {Chiaki}, \& {Yorke}}]{Hirano2014}
{Hirano}, S., {Hosokawa}, T., {Yoshida}, N., {et~al.} 2014, \apj, 781, 60

\bibitem[{{Hosokawa} {et~al.}(2015){Hosokawa}, {Hirano}, {Kuiper}, {Yorke},
  {Omukai}, \& {Yoshida}}]{Hosokawa2015}
{Hosokawa}, T., {Hirano}, S., {Kuiper}, R., {et~al.} 2015, ArXiv e-prints,
  arXiv:1510.01407

\bibitem[{{Hosokawa} {et~al.}(2011){Hosokawa}, {Omukai}, {Yoshida}, \&
  {Yorke}}]{Hosokawa2011}
{Hosokawa}, T., {Omukai}, K., {Yoshida}, N., \& {Yorke}, H.~W. 2011, Science,
  334, 1250

\bibitem[{{Howes} {et~al.}(2014){Howes}, {Asplund}, {Casey}, {Keller}, {Yong},
  {Gilmore}, {Lind}, {Worley}, {Bessell}, {Casagrande}, {Marino}, {Nataf},
  {Owen}, {Da Costa}, {Schmidt}, {Tisserand}, {Randich}, {Feltzing},
  {Vallenari}, {Allende Prieto}, {Bensby}, {Flaccomio}, {Korn}, {Pancino},
  {Recio-Blanco}, {Smiljanic}, {Bergemann}, {Costado}, {Damiani}, {Heiter},
  {Hill}, {Hourihane}, {Jofr{\'e}}, {Lardo}, {de Laverny}, {Magrini},
  {Maiorca}, {Masseron}, {Morbidelli}, {Sacco}, {Minniti}, \&
  {Zoccali}}]{Howes2014}
{Howes}, L.~M., {Asplund}, M., {Casey}, A.~R., {et~al.} 2014, \mnras, 445, 4241

\bibitem[{{Howes} {et~al.}(2015){Howes}, {Casey}, {Asplund}, {Keller}, {Yong},
  {Nataf}, {Poleski}, {Lind}, {Kobayashi}, {Owen}, {Ness}, {Bessell}, {da
  Costa}, {Schmidt}, {Tisserand}, {Udalski}, {Szyma{\'n}ski}, {Soszy{\'n}ski},
  {Pietrzy{\'n}ski}, {Ulaczyk}, {Wyrzykowski}, {Pietrukowicz}, {Skowron},
  {Koz{\l}owski}, \& {Mr{\'o}z}}]{Howes2015}
{Howes}, L.~M., {Casey}, A.~R., {Asplund}, M., {et~al.} 2015, \nat, 527, 484

\bibitem[{{Ishiyama} {et~al.}(2015){Ishiyama}, {Enoki}, {Kobayashi}, {Makiya},
  {Nagashima}, \& {Oogi}}]{Ishiyama2015}
{Ishiyama}, T., {Enoki}, M., {Kobayashi}, M.~A.~R., {et~al.} 2015, \pasj, 67,
  61

\bibitem[{{Ishiyama} {et~al.}(2009{\natexlab{a}}){Ishiyama}, {Fukushige}, \&
  {Makino}}]{Ishiyama2009b}
{Ishiyama}, T., {Fukushige}, T., \& {Makino}, J. 2009{\natexlab{a}}, \pasj, 61,
  1319

\bibitem[{{Ishiyama} {et~al.}(2009{\natexlab{b}}){Ishiyama}, {Fukushige}, \&
  {Makino}}]{Ishiyama2009}
---. 2009{\natexlab{b}}, \apj, 696, 2115

\bibitem[{Ishiyama {et~al.}(2012)Ishiyama, Nitadori, \& Makino}]{Ishiyama2012}
Ishiyama, T., Nitadori, K., \& Makino, J. 2012, in Proc. Int. Conf. High
  Performance Computing, Networking, Storage and Analysis, SC'12 (Los Alamitos,
  CA: IEEE Computer Society Press), 5:, (arXiv:1211.4406)

\bibitem[{{Keller} {et~al.}(2007){Keller}, {Schmidt}, {Bessell}, {Conroy},
  {Francis}, {Granlund}, {Kowald}, {Oates}, {Martin-Jones}, {Preston},
  {Tisserand}, {Vaccarella}, \& {Waterson}}]{Keller2007}
{Keller}, S.~C., {Schmidt}, B.~P., {Bessell}, M.~S., {et~al.} 2007,
  Publications of the Astronomical Society of Australia, 24, 1

\bibitem[{{Keller} {et~al.}(2014){Keller}, {Bessell}, {Frebel}, {Casey},
  {Asplund}, {Jacobson}, {Lind}, {Norris}, {Yong}, {Heger}, {Magic}, {da
  Costa}, {Schmidt}, \& {Tisserand}}]{Keller2014}
{Keller}, S.~C., {Bessell}, M.~S., {Frebel}, A., {et~al.} 2014, \nat, 506, 463

\bibitem[{{Kinugawa} {et~al.}(2016){Kinugawa}, {Miyamoto}, {Kanda}, \&
  {Nakamura}}]{Kinugawa2015}
{Kinugawa}, T., {Miyamoto}, A., {Kanda}, N., \& {Nakamura}, T. 2016, \mnras,
  456, 1093

\bibitem[{{Kitayama} {et~al.}(2001){Kitayama}, {Susa}, {Umemura}, \&
  {Ikeuchi}}]{Kitayama2001}
{Kitayama}, T., {Susa}, H., {Umemura}, M., \& {Ikeuchi}, S. 2001, \mnras, 326,
  1353

\bibitem[{{Klypin} {et~al.}(1999){Klypin}, {Kravtsov}, {Valenzuela}, \&
  {Prada}}]{Klypin1999}
{Klypin}, A., {Kravtsov}, A.~V., {Valenzuela}, O., \& {Prada}, F. 1999, \apj,
  522, 82

\bibitem[{{Komiya} {et~al.}(2010){Komiya}, {Habe}, {Suda}, \&
  {Fujimoto}}]{Komiya2010}
{Komiya}, Y., {Habe}, A., {Suda}, T., \& {Fujimoto}, M.~Y. 2010, \apj, 717, 542

\bibitem[{{Komiya} {et~al.}(2015){Komiya}, {Suda}, \& {Fujimoto}}]{Komiya2015}
{Komiya}, Y., {Suda}, T., \& {Fujimoto}, M.~Y. 2015, \apjl, 808, L47

\bibitem[{{Koposov} {et~al.}(2015){Koposov}, {Belokurov}, {Torrealba}, \&
  {Evans}}]{Koposov2015}
{Koposov}, S.~E., {Belokurov}, V., {Torrealba}, G., \& {Evans}, N.~W. 2015,
  \apj, 805, 130

\bibitem[{{Kroupa}(2001)}]{Kroupa2001}
{Kroupa}, P. 2001, \mnras, 322, 231

\bibitem[{{Lacey} \& {Cole}(1993)}]{Lacey1993}
{Lacey}, C., \& {Cole}, S. 1993, \mnras, 262, 627

\bibitem[{{Laevens} {et~al.}(2015){Laevens}, {Martin}, {Bernard}, {Schlafly},
  {Sesar}, {Rix}, {Bell}, {Ferguson}, {Slater}, {Sweeney}, {Wyse}, {Huxor},
  {Burgett}, {Chambers}, {Draper}, {Hodapp}, {Kaiser}, {Magnier}, {Metcalfe},
  {Tonry}, {Wainscoat}, \& {Waters}}]{Laevens2015}
{Laevens}, B.~P.~M., {Martin}, N.~F., {Bernard}, E.~J., {et~al.} 2015, \apj,
  813, 44

\bibitem[{{Lewis} {et~al.}(2000){Lewis}, {Challinor}, \& {Lasenby}}]{Lewis2000}
{Lewis}, A., {Challinor}, A., \& {Lasenby}, A. 2000, \apj, 538, 473

\bibitem[{{Li} {et~al.}(2015){Li}, {Zhao}, {Christlieb}, {Wang}, {Wang},
  {Zhang}, {Hou}, \& {Yuan}}]{Li2015}
{Li}, H.-N., {Zhao}, G., {Christlieb}, N., {et~al.} 2015, \apj, 798, 110

\bibitem[{{Luo} {et~al.}(2015){Luo}, {Zhao}, {Zhao}, {Deng}, {Liu}, {Jing},
  {Wang}, {Zhang}, {Shi}, {Cui}, {Chu}, {Li}, {Bai}, {Cai}, {Cao}, {Cao},
  {Carlin}, {Chen}, {Chen}, {Chen}, {Chen}, {Chen}, {Chen}, {Chen},
  {Christlieb}, {Chu}, {Cui}, {Dong}, {Du}, {Fan}, {Feng}, {Fu}, {Gao}, {Gong},
  {Gu}, {Guo}, {Han}, {He}, {Hou}, {Hou}, {Hou}, {Hu}, {Hu}, {Hu}, {Huo},
  {Jia}, {Jiang}, {Jiang}, {Jiang}, {Jin}, {Kong}, {Kong}, {Lei}, {Li}, {Li},
  {Li}, {Li}, {Li}, {Li}, {Li}, {Li}, {Li}, {Li}, {Li}, {Li}, {Liang}, {Lin},
  {Liu}, {Liu}, {Liu}, {Liu}, {Lu}, {Luo}, {Mao}, {Newberg}, {Ni}, {Qi}, {Qi},
  {Shen}, {Shi}, {Song}, {Song}, {Su}, {Su}, {Tang}, {Tao}, {Tian}, {Wang},
  {Wang}, {Wang}, {Wang}, {Wang}, {Wang}, {Wang}, {Wang}, {Wang}, {Wang},
  {Wang}, {Wang}, {Wang}, {Wang}, {Wang}, {Wang}, {Wang}, {Wang}, {Wang},
  {Wang}, {Wei}, {Wei}, {Wu}, {Wu}, {Wu}, {Wu}, {Wu}, {Xing}, {Xu}, {Xu}, {Xu},
  {Yan}, {Yang}, {Yang}, {Yang}, {Yang}, {Yao}, {Yu}, {Yuan}, {Yuan}, {Yuan},
  {Yuan}, {Zhai}, {Zhang}, {Zhang}, {Zhang}, {Zhang}, {Zhang}, {Zhang},
  {Zhang}, {Zhang}, {Zhao}, {Zhou}, {Zhou}, {Zhu}, {Zhu}, {Zou}, \&
  {Zuo}}]{Luo2015}
{Luo}, A.-L., {Zhao}, Y.-H., {Zhao}, G., {et~al.} 2015, ArXiv e-prints,
  arXiv:1505.01570

\bibitem[{{Machacek} {et~al.}(2001){Machacek}, {Bryan}, \&
  {Abel}}]{Machacek2001}
{Machacek}, M.~E., {Bryan}, G.~L., \& {Abel}, T. 2001, \apj, 548, 509

\bibitem[{{Machida} \& {Doi}(2013)}]{Machida2013}
{Machida}, M.~N., \& {Doi}, K. 2013, \mnras, 435, 3283

\bibitem[{{Marigo} {et~al.}(2001){Marigo}, {Girardi}, {Chiosi}, \&
  {Wood}}]{Marigo2001}
{Marigo}, P., {Girardi}, L., {Chiosi}, C., \& {Wood}, P.~R. 2001, \aap, 371,
  152

\bibitem[{{Moore} {et~al.}(1999){Moore}, {Quinn}, {Gvoernato}, {Stadel}, \&
  {Lake}}]{Moore1999}
{Moore}, B., {Quinn}, T., {Gvoernato}, F., {Stadel}, J., \& {Lake}, G. 1999,
  \mnras, 310, 1147

\bibitem[{{Nishi} \& {Susa}(1999)}]{Nishi1999}
{Nishi}, R., \& {Susa}, H. 1999, \apjl, 523, L103

\bibitem[{{Nitadori} {et~al.}(2006){Nitadori}, {Makino}, \&
  {Hut}}]{Nitadori2006}
{Nitadori}, K., {Makino}, J., \& {Hut}, P. 2006, \na, 12, 169

\bibitem[{{Okamoto} \& {Frenk}(2009)}]{Okamoto2009}
{Okamoto}, T., \& {Frenk}, C.~S. 2009, \mnras, 399, L174

\bibitem[{{Omukai} \& {Nishi}(1998)}]{Omukai1998}
{Omukai}, K., \& {Nishi}, R. 1998, \apj, 508, 141

\bibitem[{{Omukai} \& {Palla}(2001)}]{Omukai2001}
{Omukai}, K., \& {Palla}, F. 2001, \apjl, 561, L55

\bibitem[{{Omukai} \& {Palla}(2003)}]{Omukai2003}
---. 2003, \apj, 589, 677

\bibitem[{{Planck Collaboration} {et~al.}(2014){Planck Collaboration}, {Ade},
  {Aghanim}, {Armitage-Caplan}, {Arnaud}, {Ashdown}, {Atrio-Barandela},
  {Aumont}, {Baccigalupi}, {Banday}, \& et~al.}]{Planck2014}
{Planck Collaboration}, {Ade}, P.~A.~R., {Aghanim}, N., {et~al.} 2014, \aap,
  571, A16

\bibitem[{{Power} {et~al.}(2003){Power}, {Navarro}, {Jenkins}, {Frenk},
  {White}, {Springel}, {Stadel}, \& {Quinn}}]{Power2003}
{Power}, C., {Navarro}, J.~F., {Jenkins}, A., {et~al.} 2003, \mnras, 338, 14

\bibitem[{{Press} \& {Schechter}(1974)}]{Press1974}
{Press}, W.~H., \& {Schechter}, P. 1974, \apj, 187, 425

\bibitem[{{Robin} {et~al.}(2003){Robin}, {Reyl{\'e}}, {Derri{\`e}re}, \&
  {Picaud}}]{Robin2003}
{Robin}, A.~C., {Reyl{\'e}}, C., {Derri{\`e}re}, S., \& {Picaud}, S. 2003,
  \aap, 409, 523

\bibitem[{{Sakurai} {et~al.}(2016){Sakurai}, {Vorobyov}, {Hosokawa}, {Yoshida},
  {Omukai}, \& {Yorke}}]{Sakurai2015}
{Sakurai}, Y., {Vorobyov}, E.~I., {Hosokawa}, T., {et~al.} 2016, \mnras, 459,
  1137

\bibitem[{{Scannapieco} {et~al.}(2006){Scannapieco}, {Kawata}, {Brook},
  {Schneider}, {Ferrara}, \& {Gibson}}]{Scannapieco2006}
{Scannapieco}, E., {Kawata}, D., {Brook}, C.~B., {et~al.} 2006, \apj, 653, 285

\bibitem[{{Sheinis} {et~al.}(2014){Sheinis}, {Barden}, {Birchall}, {Carollo},
  {Bland-Hawthorn}, {Brzeski}, {Case}, {Cannon}, {Churilov}, {Couch}, {Dean},
  {De Silva}, {D'Orazi}, {Farrell}, {Fiegert}, {Freeman}, {Frost}, {Gers},
  {Goodwin}, {Gray}, {Heald}, {Heijmans}, {Jones}, {Keller}, {Klauser},
  {Kondrat}, {Lawrence}, {Lee}, {Mali}, {Martell}, {Mathews}, {Mayfield},
  {Miziarski}, {Muller}, {Pai}, {Patterson}, {Penny}, {Orr}, {Shortridge},
  {Simpson}, {Smedley}, {Smith}, {Stafford}, {Staszak}, {Vuong}, {Waller},
  {Wylie de Boer}, {Xavier}, {Zheng}, {Zhelem}, \& {Zucker}}]{Sheinis2014}
{Sheinis}, A., {Barden}, S., {Birchall}, M., {et~al.} 2014, in Society of
  Photo-Optical Instrumentation Engineers (SPIE) Conference Series, Vol. 9147,
  Society of Photo-Optical Instrumentation Engineers (SPIE) Conference Series,
  91470Y

\bibitem[{{Shigeyama} {et~al.}(2003){Shigeyama}, {Tsujimoto}, \&
  {Yoshii}}]{Shigeyama2003}
{Shigeyama}, T., {Tsujimoto}, T., \& {Yoshii}, Y. 2003, \apjl, 586, L57

\bibitem[{{Smith} {et~al.}(2011){Smith}, {Glover}, {Clark}, {Greif}, \&
  {Klessen}}]{Smith2011}
{Smith}, R.~J., {Glover}, S.~C.~O., {Clark}, P.~C., {Greif}, T., \& {Klessen},
  R.~S. 2011, \mnras, 414, 3633

\bibitem[{{Stacy} \& {Bromm}(2014)}]{Stacy2014}
{Stacy}, A., \& {Bromm}, V. 2014, \apj, 785, 73

\bibitem[{{Stacy} {et~al.}(2011){Stacy}, {Bromm}, \& {Loeb}}]{Stacy2011}
{Stacy}, A., {Bromm}, V., \& {Loeb}, A. 2011, \apjl, 730, L1

\bibitem[{{Stacy} {et~al.}(2012){Stacy}, {Greif}, \& {Bromm}}]{Stacy2012}
{Stacy}, A., {Greif}, T.~H., \& {Bromm}, V. 2012, \mnras, 422, 290

\bibitem[{{Strigari} {et~al.}(2008){Strigari}, {Bullock}, {Kaplinghat},
  {Simon}, {Geha}, {Willman}, \& {Walker}}]{Strigari2008}
{Strigari}, L.~E., {Bullock}, J.~S., {Kaplinghat}, M., {et~al.} 2008, \nat,
  454, 1096

\bibitem[{{Susa}(2013)}]{Susa2013}
{Susa}, H. 2013, \apj, 773, 185

\bibitem[{{Susa} {et~al.}(2014){Susa}, {Hasegawa}, \& {Tominaga}}]{Susa2014}
{Susa}, H., {Hasegawa}, K., \& {Tominaga}, N. 2014, \apj, 792, 32

\bibitem[{{Takada} {et~al.}(2014){Takada}, {Ellis}, {Chiba}, {Greene},
  {Aihara}, {Arimoto}, {Bundy}, {Cohen}, {Dor{\'e}}, {Graves}, {Gunn},
  {Heckman}, {Hirata}, {Ho}, {Kneib}, {F{\`e}vre}, {Lin}, {More}, {Murayama},
  {Nagao}, {Ouchi}, {Seiffert}, {Silverman}, {Sodr{\'e}}, {Spergel}, {Strauss},
  {Sugai}, {Suto}, {Takami}, \& {Wyse}}]{Takada2014}
{Takada}, M., {Ellis}, R.~S., {Chiba}, M., {et~al.} 2014, \pasj, 66, 1

\bibitem[{{Tanikawa} {et~al.}(2013){Tanikawa}, {Yoshikawa}, {Nitadori}, \&
  {Okamoto}}]{Tanikawa2013}
{Tanikawa}, A., {Yoshikawa}, K., {Nitadori}, K., \& {Okamoto}, T. 2013, \na,
  19, 74

\bibitem[{{Tanikawa} {et~al.}(2012){Tanikawa}, {Yoshikawa}, {Okamoto}, \&
  {Nitadori}}]{Tanikawa2012}
{Tanikawa}, A., {Yoshikawa}, K., {Okamoto}, T., \& {Nitadori}, K. 2012, \na,
  17, 82

\bibitem[{{Tegmark} {et~al.}(1997){Tegmark}, {Silk}, {Rees}, {Blanchard},
  {Abel}, \& {Palla}}]{Tegmark1997}
{Tegmark}, M., {Silk}, J., {Rees}, M.~J., {et~al.} 1997, \apj, 474, 1

\bibitem[{{Tseliakhovich} {et~al.}(2011){Tseliakhovich}, {Barkana}, \&
  {Hirata}}]{Tseliakhovich2011}
{Tseliakhovich}, D., {Barkana}, R., \& {Hirata}, C.~M. 2011, \mnras, 418, 906

\bibitem[{{Tseliakhovich} \& {Hirata}(2010)}]{Tseliakhovich2010}
{Tseliakhovich}, D., \& {Hirata}, C. 2010, \prd, 82, 083520

\bibitem[{{Tumlinson}(2010)}]{Tumlinson2010}
{Tumlinson}, J. 2010, \apj, 708, 1398

\bibitem[{{Umemura} {et~al.}(2012){Umemura}, {Susa}, {Hasegawa}, {Suwa}, \&
  {Semelin}}]{Umemura2012}
{Umemura}, M., {Susa}, H., {Hasegawa}, K., {Suwa}, T., \& {Semelin}, B. 2012,
  Progress of Theoretical and Experimental Physics, 2012, 010000

\bibitem[{{Vorobyov} \& {Basu}(2015)}]{Vorobyov2015}
{Vorobyov}, E.~I., \& {Basu}, S. 2015, \apj, 805, 115

\bibitem[{{White} \& {Springel}(2000)}]{White2000}
{White}, S.~D.~M., \& {Springel}, V. 2000, in The First Stars, ed. A.~{Weiss},
  T.~G. {Abel}, \& V.~{Hill}, 327

\bibitem[{{Yanny} {et~al.}(2009){Yanny}, {Rockosi}, {Newberg}, {Knapp},
  {Adelman-McCarthy}, {Alcorn}, {Allam}, {Allende Prieto}, {An}, {Anderson},
  {Anderson}, {Bailer-Jones}, {Bastian}, {Beers}, {Bell}, {Belokurov},
  {Bizyaev}, {Blythe}, {Bochanski}, {Boroski}, {Brinchmann}, {Brinkmann},
  {Brewington}, {Carey}, {Cudworth}, {Evans}, {Evans}, {Gates}, {G{\"a}nsicke},
  {Gillespie}, {Gilmore}, {Nebot Gomez-Moran}, {Grebel}, {Greenwell}, {Gunn},
  {Jordan}, {Jordan}, {Harding}, {Harris}, {Hendry}, {Holder}, {Ivans},
  {Ivezi{\v c}}, {Jester}, {Johnson}, {Kent}, {Kleinman}, {Kniazev},
  {Krzesinski}, {Kron}, {Kuropatkin}, {Lebedeva}, {Lee}, {French Leger},
  {L{\'e}pine}, {Levine}, {Lin}, {Long}, {Loomis}, {Lupton}, {Malanushenko},
  {Malanushenko}, {Margon}, {Martinez-Delgado}, {McGehee}, {Monet}, {Morrison},
  {Munn}, {Neilsen}, {Nitta}, {Norris}, {Oravetz}, {Owen}, {Padmanabhan},
  {Pan}, {Peterson}, {Pier}, {Platson}, {Re Fiorentin}, {Richards}, {Rix},
  {Schlegel}, {Schneider}, {Schreiber}, {Schwope}, {Sibley}, {Simmons},
  {Snedden}, {Allyn Smith}, {Stark}, {Stauffer}, {Steinmetz}, {Stoughton},
  {SubbaRao}, {Szalay}, {Szkody}, {Thakar}, {Sivarani}, {Tucker}, {Uomoto},
  {Vanden Berk}, {Vidrih}, {Wadadekar}, {Watters}, {Wilhelm}, {Wyse}, {Yarger},
  \& {Zucker}}]{Yanny2009}
{Yanny}, B., {Rockosi}, C., {Newberg}, H.~J., {et~al.} 2009, \aj, 137, 4377

\bibitem[{{Yoshida} {et~al.}(2003){Yoshida}, {Abel}, {Hernquist}, \&
  {Sugiyama}}]{Yoshida2003}
{Yoshida}, N., {Abel}, T., {Hernquist}, L., \& {Sugiyama}, N. 2003, \apj, 592,
  645

\bibitem[{{Yoshida} {et~al.}(2008){Yoshida}, {Omukai}, \&
  {Hernquist}}]{Yoshida2008}
{Yoshida}, N., {Omukai}, K., \& {Hernquist}, L. 2008, Science, 321, 669

\bibitem[{{Yoshida} {et~al.}(2006){Yoshida}, {Omukai}, {Hernquist}, \&
  {Abel}}]{Yoshida2006}
{Yoshida}, N., {Omukai}, K., {Hernquist}, L., \& {Abel}, T. 2006, \apj, 652, 6

\bibitem[{{Yoshii}(1981)}]{Yoshii1981}
{Yoshii}, Y. 1981, \aap, 97, 280

\end{thebibliography}

\end{document}